\documentclass[reprint,aps,prb,groupedaddress,floatfix]{revtex4-2}

\usepackage{amsmath,amssymb,mathtools}
\usepackage{physics}
\usepackage{graphicx}
\usepackage{enumerate}
\usepackage{float}
\usepackage{xcolor}

\usepackage[hidelinks]{hyperref}

\newcommand{\pati}[2]{}

\begin{document}

\title{Space-Time Event Scattering and Extension Method (STESEM): A Universal Framework for Scattering in Space-Time Metamaterials}

\author{Klaas De Kinder}
\author{Amir Bahrami}
\author{Christophe Caloz}
\email[]{christophe.caloz@kuleuven.be}
\affiliation{Department of Electrical Engineering, KU Leuven, Leuven, 3000, Belgium}  

\date{\today}

\begin{abstract}
    Space-time metamaterials offer unprecedented control over electromagnetic waves by enabling simultaneous manipulation of spatial and temporal degrees of freedom. However, analytical descriptions of their scattering processes remain fragmented, with existing approaches typically requiring configuration-specific derivations or transformations to specialized reference frames that become impractical for accelerated or multi-interface structures. In this tutorial, we introduce the space-time event scattering and extension method (STESEM), a universal framework for electromagnetic scattering at arbitrary space-time interfaces. By decomposing the scattering process into a local interaction event and a subsequent extension along invariant traveling-wave coordinates, STESEM formulates space-time scattering directly in the laboratory frame without requiring coordinate transformations. The method provides a unified description of scattering amplitudes, frequency transitions and phase transformations for arbitrary incident waveforms and interface trajectories. We demonstrate the framework by deriving the complete scattering responses of canonical space-time structures, including stationary and instantaneous interfaces, space-time corners, uniformly moving interfaces, wedges and accelerated boundaries. Beyond providing analytical solutions, STESEM reveals the physical principles underlying these distinct phenomena and shows that complex space-time scattering processes can be constructed from elementary moving-interface interactions. This framework establishes a systematic foundation for analyzing and designing advanced space-time metamaterials, enabling extensions toward dispersive, bianisotropic and higher-dimensional systems.
\end{abstract}

\maketitle

\section{Introduction}\label{sec:Introduction}

\pati{Metamaterials}{}

Metamaterials are artificial media composed of subwavelength yet supermolecular particles engineered to exhibit electromagnetic properties unavailable in naturally occurring materials~\cite{Walser2001_Electro_META_PUB}. They have a long history, extending from ancient composites and artificial dielectrics to modern structures with novel physical properties~\cite{Tretyakov2026_History_PUB,Cui2024_Roadmap_PUB}. Their constituent elements and spatial arrangements can be tailored to provide unprecedented control over wave propagation and enable exotic electromagnetic phenomena, including negative refraction~\cite{Veselago1968_Neg_Ref_PUB,Pendry2000_Neg_Refr_Lens_PUB}, electromagnetic cloaking~\cite{Schurig2006_Cloaking_PUB}, extreme anisotropy~\cite{Schurig2003_Indef_Eps_Mu_PUB}, artificial magnetism at high frequencies~\cite{Soukoulis2004_META_PUB}, engineered dispersion~\cite{Caloz2011_Disp_Eng_PUB,Capasso2020_Disp_Eng_Surf_PUB}, nonlocal electromagnetic responses~\cite{Silveirinha2006_Nonlocal_PUB} and magnetless nonreciprocity~\cite{Caloz2017_Nonreci_Nongyro_Magn_Metasur_PUB}. This versatility has established metamaterials as a powerful platform for realizing electromagnetic functionalities beyond those attainable with conventional materials.

\pati{Temporal Metamaterials}{}

While most metamaterial research has focused primarily on spatial structuring, the concept can be extended to the \emph{temporal domain} by dynamically varying the material parameters through temporal modulation. This new class of metamaterials was first investigated in the pioneering works of Morgenthaler~\cite{Morgenthaler1958_TEM_PUB} and Weinstein~\cite{Weinstein1965_Stretching_TEM_PUB}, which revealed the effects of time-varying media on electromagnetic waves. Shortly thereafter, temporal modulation was experimentally demonstrated in signal-processing systems~\cite{Elliott1966_Signal_Proc_TEM_CONF}. Introducing time as an additional design dimension fundamentally alters wave dynamics, giving rise to phenomena with no direct counterpart in static media. These include time refraction~\cite{Mendonca2002_Time_Refr_and_Refl_TEM_PUB}, time reversal~\cite{Fink2016_TR_TEM_PUB}, operation beyond fundamental bounds~\cite{Hadad2018_Bode-Fano_Bound_TEM_PUB,Alu2019_Chu_Limit_TEM_PUB,Monticone2024_Rozanov_Bound_TEM_PUB}, inverse-prism decomposition~\cite{Caloz2018_Inverse_Prism_TEM_PUB}, temporal impedance matching~\cite{Engheta2020_Aiming_TEM_PUB}, temporal aiming~\cite{Engheta2020_Coating_TEM_PUB}, beam splitting~\cite{Guerreiro2003_Split_Interf_TEM_PUB}, photon generation~\cite{Guerreiro2000_PUB}, polarization conversion~\cite{Werner2021_PUB} and temporal analogues of Faraday rotation~\cite{Alu2022_Nonreci_Faraday_TEM_PUB,Huanan2023_Faraday_TEM_PUB}.

\pati{Space-Time Metamaterials}{}

Combining spatial structuring with temporal modulation leads to the broader class of \emph{space-time} metamaterials~\cite{Caloz2019a_ST_Metamaterials_USTEM_PUB,Caloz2019b_ST_Metamaterials_USTEM_PUB,Caloz2022_GSTEMs_ASTEM_PUB,Luo2026_Review_USTEM}, which were first investigated in the seminal works of Cassedy and Oliner~\cite{Cassedy1963_PUB,Cassedy1967_PUB}. In these media, the material properties vary simultaneously in space and time, coupling spatial and temporal wave transformations within a single platform and giving rise to phenomena inaccessible to either purely spatial or purely temporal modulation. These include space-time reversal~\cite{Deck-Leger2018_Refocus_USTEM_PUB}, Fresnel-Fizeau pseudo-drag~\cite{Pendry2019_Fresnel_Drag_USTEM_PUB,Pendry2021_Homogenization_USTEM_PUB,Qiu2022_Diff_Fizeau_USTEM_PUB}, gravitational analogues~\cite{Bahrami2023_ASTEMs_PUB}, Doppler-based pulse amplification~\cite{DeKinder2025_DoPA_USTEM_PUB}, dynamic diffraction~\cite{Eleftheriades2019_Diff_Grat_USTEM_PUB}, photon cooling~\cite{Pendry2024_Air_Cond_Phot_USTEM_PUB}, generalized frequency chirping at accelerated interfaces~\cite{DeKinder2026_Scat_Chirp_ASTEM_PUB} or gradient-index interfaces~\cite{Li2025_Graded_Index_USTEM_PUB} and arbitrary pulse shaping~\cite{Bahrami2025_Pulse_Shap_ASTEM_PUB}.

\pati{Interface}{}

Space-time metamaterials may be viewed as dimensional and functional generalizations of acousto-optic modulators~\cite{Saleh2019_BOOK}, in which the modulation of one or more medium parameters, such as the refractive index, undergoes \emph{synthetic motion}. These metamaterials can be interpreted as sequences of moving \emph{interfaces}, which represent the metaparticles or ``metamolecules'' of the material and whose individual scattering responses collectively govern the behavior of the overall structure. Beyond serving as the fundamental building blocks of space-time metamaterials, these interfaces exhibit rich physics in their own right, most notably through the simultaneous transformation of frequency and momentum~\cite{Deck-Leger2019_Uni_Vel_USTEM_PUB}. They may then occur in isolation or be assembled into elementary configurations such as corners~\cite{Alu2025_ST_Corner_USTEM_PUB}, wedges~\cite{Bahrami2025_Wedges_USTEM_PUB} or cavities~\cite{DeKinder2026_Scat_Chirp_ASTEM_PUB}, giving rise to multiple scattering processes involving successive Doppler shifts, wave conversions and phase accumulations. Because a space-time metamaterial is assembled from such interfaces, much of its essential physics is already encoded in their individual scattering responses, while the behavior of the complete structure can subsequently be constructed using standard techniques such as Bloch--Floquet analysis and transfer-matrix methods. For this reason, we restrict our attention in this paper to space-time interfaces.

\pati{Gap}{}

Existing analytical treatments of space-time interfaces have predominantly been developed on a case-by-case basis, with each configuration requiring a dedicated formulation. Most approaches rely on coordinate transformations to a comoving frame~\cite{Lampe1978_Inter_EM_USTEM_PUB,Pendry2019_Fresnel_Drag_USTEM_PUB,Pendry2021_Homogenization_USTEM_PUB,Caloz2019b_ST_Metamaterials_USTEM_PUB,Deck-Leger2019_Uni_Vel_USTEM_PUB,Bahrami2023_ASTEMs_PUB}, implemented through Lorentz or Rindler transformations. Although such transformations yield elegant solutions for constant- and unique-velocity configurations, they introduce fundamental limitations and become cumbersome or inapplicable in the case of accelerated interfaces, multiple interfaces moving at different velocities or arbitrary combinations of space-time interfaces, for which a single comoving frame generally does not exist. There is currently no general theoretical framework capable of describing arbitrary space-time scattering processes in a unified, systematic and physically transparent manner.

\pati{Contribution}{}

Here we introduce the \emph{space-time event scattering and extension method (STESEM)} as a \emph{general framework} for analyzing wave interactions with general structures composed of arbitrary space-time interfaces~\cite{Bahrami2025_Retard_Arg_ASTEM_CONF}. Unlike previous approaches, which treat individual configurations separately, STESEM establishes a unified methodology based on the intrinsic properties of traveling waves and elementary space-time scattering events. The method eliminates the need for geometry-dependent coordinate transformations by formulating the problem directly in the laboratory frame. It therefore naturally extends beyond constant- and unique-velocity interfaces to accelerated space-time boundaries and configurations involving multiple interfaces moving at different velocities. Moreover, STESEM provides the complete scattering response for an arbitrary incident waveform within a single formulation, including the scattering coefficients, frequency transitions and phase shifts that traditionally require separate derivations. The framework consequently enables the systematic analysis of space-time interfaces, wedges, corners and more complex structures as combinations of elementary space-time scattering elements. This tutorial develops STESEM progressively, providing both physical insight and practical analytical tools for studying a broad class of space-time-varying metamaterials.

\pati{Paper Structure}{}

The paper is organized as follows. Section~\ref{sec:Statement_of_the_Problem} formulates the general space-time scattering problem. Section~\ref{sec:STESEM} then introduces the STESEM concept and its underlying principles. Next, Sec.~\ref{sec:Canonical_Space-Time_Structures} presents the canonical space-time structures considered throughout the paper, while Sec.~\ref{sec:Applications} derives their scattering responses using the STESEM framework. Finally, Sec.~\ref{sec:Conclusions} summarizes the main results and discusses possible extensions of STESEM to more complex space-time metamaterial systems.

%%%%%%%%%%%%%%%%%%%%%%%%%%%%%%%%%%%%%%%%%%%%%%%%%%%%%%%%%%%%%%%%%%%%%%%%%%%%%%%%%%%%%%%%%%%%%%%%%%%%

\section{Statement of the Problem}\label{sec:Statement_of_the_Problem}

\pati{Space-Time Metamaterial and Its Implementation}{}

\begin{figure}[h!]
    \centering
    \includegraphics[width=1.0\linewidth]{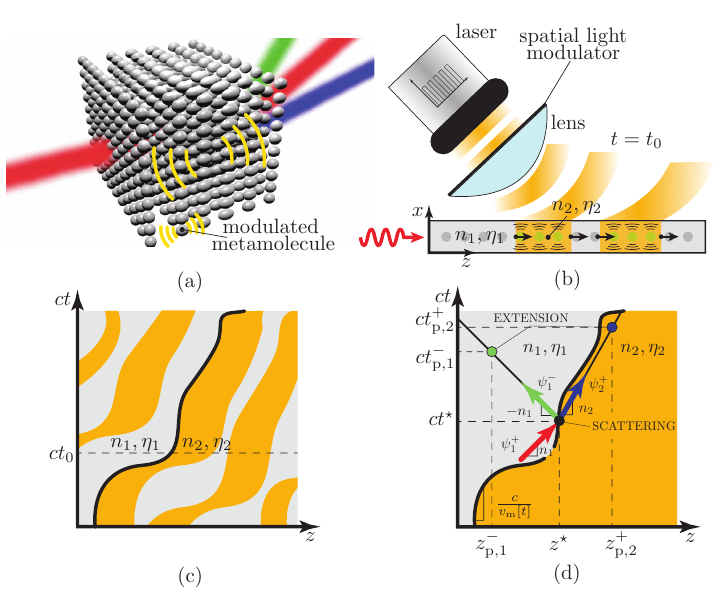}
    \caption{Overview of a space-time metamaterial structure and proposed STESEM framework. (a)~Space-time metamaterial, composed of dynamically modulated metamolecules. (b)~Conceptual ``optical guillotine'' implementation~\cite{Caloz2019a_ST_Metamaterials_USTEM_PUB}, in which an obliquely incident laser beam,
    shaped by a spatial light modulator and/or lens, induces a traveling modulation that creates a moving refractive-index and impedance profile consisting of metamolecule layers with arbitrarily engineered durations and spacings. (c)~Representation of the metamaterial in the space-time (Minkowski) diagram and identification of a specific interface (black trajectory) between adjacent medium regions. (d)~Reduction of (c) to a space-time metaatom, consisting of a single moving interface separating two media, and STESEM concept, in which the scattered fields are first determined at the moving interface by applying the related moving boundary conditions at the scattering event and then extended along the scattered-wave trajectories to construct the complete space-time field.}
    \label{fig:Problem_statement_and_STEM}
\end{figure}

Figure~\ref{fig:Problem_statement_and_STEM} provides an overview of a space-time metamaterial structure and the proposed STESEM framework. Figure~\ref{fig:Problem_statement_and_STEM}a depicts a space-time metamaterial, which is composed of metamolecules that are dynamically modulated in both space and time without any net motion of matter. By combining temporal modulation with spatial structuring, such a metamaterial produces distinctive effects, such as the simultaneous generation of novel spatial and temporal frequency components, leading to generalized refraction and diffraction~\cite{Caloz2019a_ST_Metamaterials_USTEM_PUB,Caloz2019b_ST_Metamaterials_USTEM_PUB}. Figure~\ref{fig:Problem_statement_and_STEM}b presents a conceptual optical implementation of a space-time metamaterial. An external excitation dynamically alters the electromagnetic properties of a substrate, inducing a traveling-wave modulation of its material parameters and thereby creating moving regions with different refractive indices, impedances or both. The boundaries separating these regions form space-time interfaces that constitute the elementary scattering elements, or ``metaatoms'', of the metamaterial~\footnote{Since the actual atoms and molecules of the structure only \emph{oscillate} under the external excitation (Fig.~\ref{fig:Problem_statement_and_STEM}b), with no net transport of matter, the \emph{interfaces} are the only entities exhibiting effective translational motion (Fig.~\ref{fig:Problem_statement_and_STEM}c) and may therefore be regarded as the real metaatoms of the metamaterial (Fig.~\ref{fig:Problem_statement_and_STEM}d).}. Moreover, the use of a free-form lens and/or spatial light modulator allows the durations, spacings and trajectories of these interfaces to be engineered arbitrarily.

\pati{Reduction to the Meta-atom Interface}{}

Figure~\ref{fig:Problem_statement_and_STEM}c represents the structure of Fig.~\ref{fig:Problem_statement_and_STEM}b in the space-time---or Minkowski---diagram. The curved trajectories in this diagram indicate that the modulation interfaces move dynamically and may undergo acceleration. As mentioned in Sec.~\ref{sec:Introduction}, although a complete space-time metamaterial comprises many interfaces, its fundamental scattering properties can be inferred by reducing the structure to a single space-time interface, as highlighted by the black trajectory in the figure. Such moving interfaces constitute the metaatoms of the space-time metamaterial and provide a general basis for analyzing their electromagnetic response.

\pati{System Description}{}

Figure~\ref{fig:Problem_statement_and_STEM}d shows the reduction of the metamaterial in Fig.~\ref{fig:Problem_statement_and_STEM}c to its interface metaatom. Specifically, we consider a moving abrupt~\footnote{In practice, interfaces cannot be perfectly abrupt, neither in space nor in time, and must therefore be smooth. However, if the transition region is much smaller than the wavelength for a spatial interface~\cite{Golik2022_Gradient_Refr_SEM_PUB}, much shorter than the period for a temporal interface~\cite{Pena2025_Time_Refr_Smooth_TEM_PUB} or satisfies both conditions for a space-time interface~\cite{Li2025_Graded_Index_USTEM_PUB}, the resulting scattering is indistinguishable from that produced by an abrupt interface. Abrupt interfaces therefore provide excellent approximations for subwavelength and subperiod transitions and capture their essential physical characteristics.} interface separating two isotropic, linear and nondispersive media in a one-dimensional spatial plus one-dimensional temporal (1+1D) configuration. The media are characterized by refractive indices~$n_{i}$ and impedances~$\eta_{i}$, where~$i=1,2$ labels the medium. The interface moves along the~$z$-direction and follows the trajectory~$z{\left[t\right]}$, with instantaneous velocity~$v_{\text{m}}{\left[t\right]}=\dd{z{\left[t\right]}}/\dd{t}$. The electromagnetic fields depend solely on~$z$ and~$t$, with the electric field polarized along the~$x$-direction and the magnetic field along the~$y$-direction. As illustrated in the figure, an incident wave interacts with the moving interface and generates scattered waves in both media. The objective is to determine these scattered fields for an arbitrary interface trajectory. The fields of the forward- and backward-propagating waveforms in each medium may be generally expressed as
\begin{subequations}\label{eq:General_Traveling_Waveforms}
    \begin{align}
        E_{i}^{\pm} &= \psi_{i}^{\pm}{\left[\tau_{i}^{\pm}\right]}\,, &
        H_{i}^{\pm} &= \pm \frac{1}{\eta_{i}}\psi_{i}^{\pm}{\left[\tau_{i}^{\pm}\right]}\,, \\
        D_{i}^{\pm} &= \frac{n_{i}}{c\eta_{i}}\psi_{i}^{\pm}{\left[\tau_{i}^{\pm}\right]}\,, &
        B_{i}^{\pm} &= \pm \frac{n_{i}}{c}\psi_{i}^{\pm}{\left[\tau_{i}^{\pm}\right]}\,,
    \end{align}
\end{subequations}
where the square brackets,~$\left[\cdot\right]$, delimit the arguments of the wave functions~$\psi_{i}^{\pm}$, which are themselves arbitrary~\footnote{\label{fn:waveform} The waveform function, $\psi_{i}^{\pm}$, in Eq.~\eqref{eq:General_Traveling_Waveforms} is arbitrary. For instance, for a propagating Gaussian-modulated pulse, it would be written as
\begin{equation*}\label{eq:Example_Gaussian_Modulated_Pulse}
    \psi_{i}^{\pm}{\left[n_{i}\frac{z}{c}\mp t\right]}
    =
    \exp\left(i\omega\left(n_{i}\frac{z}{c}\mp t\right)\right)
    \exp\left(-\frac{\left(n_{i}z/c\mp t\right)^{2}}{\sigma^{2}}\right)\,,
\end{equation*}
with~$\omega$ being the center frequency and~$\sigma$ the pulse width, reducing to a simple plane wave as $\sigma\rightarrow\infty$. Other common pulse shapes include hyperbolic-secant, Lorentzian, exponential, rectangular, sinc and chirped waveforms.}. The corresponding traveling-wave coordinates are defined as
\begin{equation}\label{eq:Traveling_Wave_Coordinates}
    \tau_{i}^{\pm}=n_{i}\frac{z}{c}\mp t,
\end{equation}
where the superscript~$\pm$ denotes forward~($+$) or backward~($-$) propagation and~$c$ is the speed of light in vacuum. They represent the general solutions of the wave equation and hence of Maxwell’s equations, corresponding to the arbitrary waveforms~$\psi_{i}^{\pm}$.

%%%%%%%%%%%%%%%%%%%%%%%%%%%%%%%%%%%%%%%%%%%%%%%%%%%%%%%%%%%%%%%%%%%%%%%%%%%%%%%%%%%%%%%%%%%%%%%%%%%%

\section{STESEM}\label{sec:STESEM}

\pati{Conventional Approach}{}

Existing analytical methods for space-time scattering typically treat each configuration independently, requiring a dedicated derivation or specialized analytical framework. A common strategy, often referred to as frame hopping~\cite{Bladel2012_BOOK}, consists of transforming the problem to the comoving reference frame of the interface, in which the boundary appears stationary and conventional boundary conditions can be applied. Although this approach is well suited to interfaces moving at constant velocity or constant proper acceleration, it cannot be directly extended to arbitrary accelerated interfaces or structures containing multiple interfaces moving at different velocities, since no single comoving reference frame generally exists. Moreover, the scattered frequencies, amplitudes and phase shifts are often determined through separate analytical procedures, making the overall analysis increasingly cumbersome. STESEM overcomes these limitations by solving the scattering problem directly in the \emph{laboratory frame}, thereby eliminating the need for coordinate transformations, and provides in one shot all the scattered information.

\pati{STESEM Concept}{}

The central idea of STESEM is to decompose the scattering problem of Fig.~\ref{fig:Problem_statement_and_STEM}d into two complementary steps: (i)~determine the scattered fields locally at the moving interface by enforcing the moving boundary conditions at the scattering event and (ii)~extend these fields to arbitrary space-time points by exploiting the invariance of the traveling-wave coordinate~$\left(\tau_{i}^{\pm}\right)$ along the scattered-wave trajectories. This decomposition separates the local scattering event from the subsequent wave propagation, thereby considerably simplifying the analysis of arbitrary moving interfaces. Consider an arbitrary observation point~$\left(z_{i}^{\pm},ct_{i}^{\pm}\right)$ in space-time, such as the two points~$\left(z_{1}^{-},ct_{1}^{-}\right)$ and~$\left(z_{2}^{+},ct_{2}^{+}\right)$ shown in the figure. The field observed at such a point originates from a unique scattering event at the moving interface. This event is characterized by the scattering time~$t^{\star}$ and position~$z^{\star}$, corresponding to the intersection of the wave trajectory passing through the observation point with the interface trajectory~$z{\left[t\right]}$. In the first step of STESEM, the moving boundary conditions are enforced at this scattering event~$\left(z^{\star},ct^{\star}\right)$ to determine the scattered fields at the interface. In the second step, these interface solutions are extended to the observation point by exploiting the invariance of the traveling-wave coordinate along each scattered-wave trajectory, due to the isotropic, linear and nondispersive nature of the media sandwiching the interface. Each observation point in the space-time of interest can therefore be placed in one-to-one correspondence with its associated scattering event, allowing the complete space-time field distribution to be constructed from the interface solution alone. The same two-step procedure applies independently of the interface trajectory, making STESEM applicable to stationary, uniformly moving, accelerated and more general space-time configurations without requiring coordinate transformations.

\pati{Boundary Conditions}{}

The first step of STESEM consists of determining the scattered fields at the moving interface by enforcing the electromagnetic moving boundary conditions. Evaluated at the interface position~$z{\left[t\right]}$, these conditions are given by~\cite{Pauli1981_Relativity_BOOK,Kong2008_Wave_Theory_BOOK,Caloz2019b_ST_Metamaterials_USTEM_PUB}
\begin{subequations}\label{eq:General_Boundary_Conditions}
    \begin{align}
        \left.E_{1}^{\pm}-v_{\text{m}}{\left[t^{\star}\right]}B_{1}^{\pm}\right|_{z=z{\left[t^{\star}\right]}}
        &=
        \left.E_{2}^{\pm}-v_{\text{m}}{\left[t^{\star}\right]}B_{2}^{\pm}\right|_{z=z{\left[t^{\star}\right]}} \,, \\
        \left.H_{1}^{\pm}-v_{\text{m}}{\left[t^{\star}\right]}D_{1}^{\pm}\right|_{z=z{\left[t^{\star}\right]}}
        &=
        \left.H_{2}^{\pm}-v_{\text{m}}{\left[t^{\star}\right]}D_{2}^{\pm}\right|_{z=z{\left[t^{\star}\right]}} \,.
    \end{align}
\end{subequations}
Substituting Eqs.~\eqref{eq:General_Traveling_Waveforms} into Eqs.~\eqref{eq:General_Boundary_Conditions} and inserting Eqs.~\eqref{eq:Traveling_Wave_Coordinates} into the resulting relation yields
\begin{widetext}
    \begin{subequations}\label{eq:Boundary_Conditions_Psi}
        \begin{align}
            \left(1 \mp n_{1}\frac{v_{\text{m}}{\left[t^{\star}\right]}}{c}\right)
            \psi_{1}^{\pm}{\left[n_{1}\frac{z{\left[t^{\star}\right]}}{c}\mp t^{\star}\right]}
            &=
            \left(1 \mp n_{2}\frac{v_{\text{m}}{\left[t^{\star}\right]}}{c}\right)
            \psi_{2}^{\pm}{\left[n_{2}\frac{z{\left[t^{\star}\right]}}{c}\mp t^{\star}\right]} \,, \\
            \pm\frac{1}{\eta_{1}}
            \left(1 \mp n_{1}\frac{v_{\text{m}}{\left[t^{\star}\right]}}{c}\right)
            \psi_{1}^{\pm}{\left[n_{1}\frac{z{\left[t^{\star}\right]}}{c}\mp t^{\star}\right]}
            &=
            \pm\frac{1}{\eta_{2}}
            \left(1 \mp n_{2}\frac{v_{\text{m}}{\left[t^{\star}\right]}}{c}\right)
            \psi_{2}^{\pm}{\left[n_{2}\frac{z{\left[t^{\star}\right]}}{c}\mp t^{\star}\right]} \,.
        \end{align}
    \end{subequations}
\end{widetext}
For each canonical scattering configuration, the total field on either side of the interface is obtained by summing the relevant nonzero wave components. These equations constitute the general moving boundary conditions for arbitrary interface trajectories and form the starting point for all subsequent analyses. Although they may initially appear abstract, their physical meaning and practical use will become clear through the derivations in Sec.~\ref{sec:Applications}.

%%%%%%%%%%%%%%%%%%%%%%%%%%%%%%%%%%%%%%%%%%%%%%%%%%%%%%%%%%%%%%%%%%%%%%%%%%%%%%%%%%%%%%%%%%%%%%%%%%%%

\section{Canonical Space-Time Structures}\label{sec:Canonical_Space-Time_Structures}

\pati{Classification}{}

\begin{figure*}
    \centering
    \includegraphics[width=1.0\linewidth]{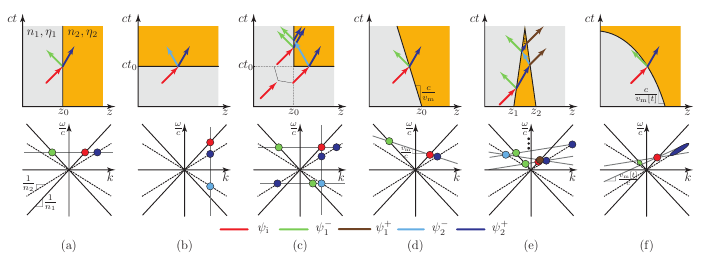}
    \caption{Canonical space-time interfaces~\cite{Caloz2019b_ST_Metamaterials_USTEM_PUB,Caloz2025_STEM_Elements_CONF}. Top row: space-time trajectory diagrams in the direct space-time domain, $(z,ct)$. Bottom row: corresponding spectral transition diagrams in the inverse (Fourier) space-time domain, $(k,\omega/c)$. (a)~Space (stationary) interface~\cite{Fresnel1834_Original_BOOK,Lorentz1875_Original_BOOK}. (b)~Time (instantaneous) interface~\cite{Morgenthaler1958_TEM_PUB,Mendonca2002_Time_Refr_and_Refl_TEM_PUB}. (c)~Space-time corner~\cite{Kalluri1988_Corner_TEM_PUB,Alu2025_ST_Corner_USTEM_PUB}. (d)~Constant-velocity (contradirectional and subluminal) traveling interface~\cite{Lampert1956_PUB,Deck-Leger2019_Uni_Vel_USTEM_PUB}. (e)~Space-time (closing) wedge~\cite{Bahrami2025_Wedges_USTEM_PUB}. (f)~Accelerated interface~\cite{DeKinder2026_Scat_Chirp_ASTEM_PUB}.}
    \label{fig:Canonical_Space-Time_Structures}
\end{figure*}

Figure~\ref{fig:Canonical_Space-Time_Structures} presents a set of canonical interfaces that form the fundamental building bricks of most space-time metamaterials (Fig.~\ref{fig:Problem_statement_and_STEM}c). Each configuration corresponds to a distinct type of space-time interface, characterized by specific scattering mechanisms and conservation properties~\cite{Caloz2019b_ST_Metamaterials_USTEM_PUB}. We shall systematically apply STESEM to all these structures in Sec.~\ref{sec:Applications}.

\pati{Canonical Configurations}{}

Figure~\ref{fig:Canonical_Space-Time_Structures}a shows the conventional space interface separating two homogeneous media~\cite{Fresnel1834_Original_BOOK,Lorentz1875_Original_BOOK}. This configuration represents the classical Fresnel scattering problem and provides the simplest reference case for the development of STESEM. Because the interface is stationary, the scattered waves preserve their frequency ($\omega$) while their momentum ($k$) changes across the interface. Figure~\ref{fig:Canonical_Space-Time_Structures}b illustrates a time interface, corresponding to an abrupt change in the material parameters occurring instantaneously throughout space~\cite{Morgenthaler1958_TEM_PUB,Mendonca2002_Time_Refr_and_Refl_TEM_PUB}. In contrast to a space interface, a time discontinuity conserves wave momentum while altering the frequency content. Figure~\ref{fig:Canonical_Space-Time_Structures}c represents a space-time corner, which is simplest configuration that combines both spatial and temporal discontinuities~\cite{Kalluri1988_Corner_TEM_PUB,Alu2025_ST_Corner_USTEM_PUB}. In this configuration, different portions of the incident wave encounter different discontinuities, resulting in cascaded scattering processes involving both momentum and frequency transformations. Figure~\ref{fig:Canonical_Space-Time_Structures}d depicts a uniformly moving interface separating two media~\cite{Lampert1956_PUB,Deck-Leger2019_Uni_Vel_USTEM_PUB}, which is the first genuinely space-time scattering configuration. The boundary propagates at a constant velocity and appears as a tilted trajectory in the space-time diagram, with a slope equal to the inverse of the interface velocity. Such an interface simultaneously modifies the frequency and momentum of the scattered waves through the Doppler effect. The figure shows a subluminal configuration, with reflected and transmitted waves. The configuration can also be superluminal, in which case the scattered wave ansatz changes to later-backward and later-forward waves~\cite{Caloz2019b_ST_Metamaterials_USTEM_PUB}, but the STESEM procedure remains otherwise identical. Figure~\ref{fig:Canonical_Space-Time_Structures}e illustrates a space-time wedge formed by two intersecting constant moving interfaces~\cite{Bahrami2025_Wedges_USTEM_PUB}. Multiple successive interactions with these moving boundaries can produce in specific geometries repeated Doppler scattering, resulting in wave trapping and an infinite sequence of scattered waves. The figure shows a closing wedge, with apex at the latest time; the wedge can also be opening, with apex at the earliest time. Finally, Fig.~\ref{fig:Canonical_Space-Time_Structures}f shows an accelerated interface whose modulation velocity varies with time~\cite{DeKinder2026_Scat_Chirp_ASTEM_PUB}. Unlike the uniformly moving interface, the accelerated interface produces a continuously varying Doppler shift, resulting in frequency chirping of the scattered waves.

\pati{Spectral Transition Diagrams}{}

The spectral representations at the bottom of the figure provide graphical constructions helping to determine the frequencies and momenta of the scattered waves. These diagrams are constructed in the momentum--frequency $\left(k,\omega/c\right)$ plane, where the dispersion relations of the two media are represented by the curves $k^{2}=n_{i}^{2}\omega^{2}/c^{2}$, with $i=1,2$. To determine the scattered states, one first identifies the point on the dispersion curve of the first medium corresponding to the incident wave, indicated by a red marker. A transition line is then drawn through this point, with a slope determined by the interface modulation velocity, as imposed by phase continuity at the interface~\cite{Gaafar2019_PUB,DeKinder2026_Scat_ST_Int_Disp_USTEM}. The intersections of this line with the dispersion curves identify the admissible scattered waves, while their coordinates directly provide the corresponding frequencies and momenta. The orientation of the transition line also reveals the conserved quantity: a horizontal transition line corresponds to a stationary interface and therefore expresses frequency conservation, whereas a vertical line corresponds to a purely temporal interface and expresses momentum conservation. An oblique line represents a constant moving interface and indicates simultaneous changes in frequency and momentum. Structures containing multiple interfaces require multiple transition lines, each associated with a distinct scattering event, so that successive intersections describe multiple scattering processes, such as the space-time corner (Fig.~\ref{fig:Canonical_Space-Time_Structures}c) and wedge (Fig.~\ref{fig:Canonical_Space-Time_Structures}d). Finally, an accelerated interface is represented by a continuous family of transition lines associated with its continuously varying velocity, producing a continuous evolution of the scattered frequencies and hence frequency chirping.

%%%%%%%%%%%%%%%%%%%%%%%%%%%%%%%%%%%%%%%%%%%%%%%%%%%%%%%%%%%%%%%%%%%%%%%%%%%%%%%%%%%%%%%%%%%%%%%%%%%%
\section{Applications}\label{sec:Applications}

\pati{Introduction 1}{}

\begin{figure*}
    \centering
    \includegraphics[width=1.0\linewidth]{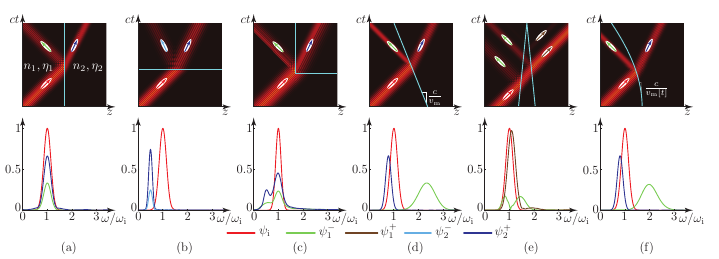}
    \caption{Scattering results obtained by STESEM for the canonical space-time structures in Fig.~\ref{fig:Canonical_Space-Time_Structures} with parameters $\left(n_{1},\eta_{1}\right)=\left(1,1\right)$ and $\left(n_{2},\eta_{2}\right)=\left(2,0.5\right)$ under a Gaussian-modulated pulse illumination. Top row: incident and scattered waves in space-time. Bottom row: corresponding temporal spectra, taken at positions away from the interface. (a)~Space interface. (b)~Time interface. (c)~Space-time corner. (d)~Constant-velocity traveling interface. (e)~Space-time wedge. (f)~Accelerated interface.}
    \label{fig:Examples_Space-Time_Structures}
\end{figure*}

We now apply STESEM, in tutorial progression, to the canonical space-time structures introduced in Fig.~\ref{fig:Canonical_Space-Time_Structures}, using the principles and general equations established in Sec.~\ref{sec:STESEM}. Figure~\ref{fig:Examples_Space-Time_Structures} presents the corresponding scattering responses to a Gaussian-modulated pulse, with the top row showing the evolution of the incident and scattered waves in space-time and the bottom row plotting the associated temporal Fourier spectra, evaluated at positions sufficiently far from the interfaces~\footnote{The apparent breaking of symmetry between space and time, arising from the restriction of the bottom diagrams to temporal frequencies, results from the~1+1D and nondispersive-medium assumptions adopted in this paper (Sec.~\ref{sec:Statement_of_the_Problem}), under which the wavevector~$\boldsymbol{k}$ reduces to the scalar quantity~$k$, linearly related to frequency by~$k=\pm n\omega/c$. Extending the analysis to~2+1D or~3+1D problems entails greater mathematical complexity, including the generalization of the fields in Eqs.~\eqref{eq:General_Traveling_Waveforms} to vectorial forms with additional components, but such problems can be addressed using the same methodology.}. The canonical structures in Fig.~\ref{fig:Canonical_Space-Time_Structures} are considered in order of increasing complexity. The STESEM analysis of each structure builds on the preceding cases, with the main text presenting only the essential analytical results and key physical interpretations, while the complete derivations are provided in the supplementary material.

\subsection{Space Interface}
\pati{Method}{}

We first consider the canonical problem of a stationary space interface (Fig.~\ref{fig:Canonical_Space-Time_Structures}a). An incident wave,~$\psi_{\text{i}}$, generates~$\psi_{1}^{-}$ (reflection) and~$\psi_{2}^{+}$ (transmission). Since the interface is stationary, its velocity is~$v_{\text{m}} = 0$ and its trajectory is simply described by~$z=z_{0}$. Consequently, the moving-interface boundary conditions [Eqs.~\eqref{eq:General_Boundary_Conditions}] reduce to the conventional continuity of the tangential electric and magnetic fields. Setting $v_{\text{m}}=0$ and substituting $z{\left[t^{\star}\right]} = z_{0}$ into Eq.~\eqref{eq:Boundary_Conditions_Psi} gives
\begin{subequations}\label{eq:Space_Interface_Boundary_Conditions}
    \begin{align}
        \psi_{\text{i}}{\left[n_{1}\frac{z_{0}}{c} - t^{\star}\right]} + \psi_{1}^{-}{\left[n_{1}\frac{z_{0}}{c} + t^{\star}\right]} &= \psi_{2}^{+}{\left[n_{2}\frac{z_{0}}{c} - t^{\star}\right]} \,, \\
        \frac{1}{\eta_{1}}\psi_{\text{i}}{\left[n_{1}\frac{z_{0}}{c} - t^{\star}\right]} - \frac{1}{\eta_{1}}\psi_{1}^{-}{\left[n_{1}\frac{z_{0}}{c} + t^{\star}\right]} &= \frac{1}{\eta_{2}}\psi_{2}^{+}{\left[n_{2}\frac{z_{0}}{c} - t^{\star}\right]} \,,
        \end{align}
\end{subequations}
Solving these equations for $\psi_{1}^{-}$ and $\psi_{2}^{+}$ then yields
\begin{subequations}\label{eq:Space_Interface_Solution_at_Interface}
    \begin{align}
        \psi_{1}^{-}{\left[n_{1}\frac{z_{0}}{c} + t^{\star}\right]} &= \frac{\eta_{2}-\eta_{1}}{\eta_{2}+\eta_{1}}\psi_{\text{i}}{\left[n_{1}\frac{z_{0}}{c} - t^{\star}\right]} \,, \\
        \psi_{2}^{+}{\left[n_{2}\frac{z_{0}}{c} - t^{\star}\right]} &= \frac{2\eta_{2}}{\eta_{2}+\eta_{1}}\psi_{\text{i}}{\left[n_{1}\frac{z_{0}}{c} - t^{\star}\right]} \,.
    \end{align}
\end{subequations}
Equations~\eqref{eq:Space_Interface_Solution_at_Interface} determine the scattered waveforms only at the scattering event,~$\left(z^{\star},ct^{\star}\right)$ (Fig.~\ref{fig:Problem_statement_and_STEM}d). 

The second task is to extend these interface solutions to arbitrary observation points in space and time. After scattering, each of the two waves propagates independently through its corresponding homogeneous medium. Since the propagation velocity is constant, every feature of the waveform (for example, the peak of a pulse) simply translates through space without changing shape. Consequently, the traveling-wave coordinate in Eq.~\eqref{eq:Traveling_Wave_Coordinates} remains invariant along the wave trajectory and uniquely labels each point of the waveform. Consider two observation points,~$\left(z_{i}^{\pm},ct_{i}^{\pm}\right)$, shown in Fig.~\ref{fig:Problem_statement_and_STEM}d. The corresponding scattering event is obtained by tracing the wave trajectory backward until it intersects the interface trajectory at~$\left(z^{\star},ct^{\star}\right)$. The trajectory of a wave passing through this point is
\begin{equation}\label{eq:Space_Interface_Trajectory_Pulse}
    z = \pm \frac{c}{n_{i}}\left(t-t_{i}^{\pm}\right)+z_{i}^{\pm} = \pm \frac{c}{n_{i}}t + \frac{c}{n_{i}}\tau_{i}^{\pm} \,,
\end{equation}
where we have denoted the traveling-wave coordinates of the observation point as~$\tau_{i}^{\pm} = n_{i}z_{i}^{\pm}/c\mp t_{i}^{\pm}$ [Eq.~\eqref{eq:Traveling_Wave_Coordinates}]. Intersecting Eq.~\eqref{eq:Space_Interface_Trajectory_Pulse} with the stationary interface position~$z_{0}$ gives the scattering time associated with the observation point,
\begin{equation}\label{eq:Space_Interface_Coordinate_Transformation}
    t^{\star} = \pm\left(n_{i}\frac{z_{0}}{c}-\tau_{i}^{\pm}\right) \,.
\end{equation}
Equation~\eqref{eq:Space_Interface_Coordinate_Transformation} establishes a one-to-one correspondence between every observation point and its associated scattering event. It essentially states that the field at any observation point is completely determined by its associated local scattering event at the interface. Substituting this equation into Eqs.~\eqref{eq:Space_Interface_Solution_at_Interface} provides the complete scattered fields,
\begin{subequations}\label{eq:Space_Interface_Scattered_Waves}
    \begin{align}
        \psi_{1}^{-}{\left[\tau_{1}^{-}\right]} &= \frac{\eta_{2}-\eta_{1}}{\eta_{2}+\eta_{1}}\psi_{\text{i}}{\left[-\tau_{1}^{-}+2n_{1}\frac{z_{0}}{c}\right]} \,, \\
        \psi_{2}^{+}{\left[\tau_{2}^{+}\right]} &= \frac{2\eta_{2}}{\eta_{2}+\eta_{1}}\psi_{\text{i}}{\left[\tau_{2}^{+} + \left(n_{1}-n_{2}\right)\frac{z_{0}}{c}\right]} \,.
    \end{align}
\end{subequations}

\pati{Two Notes}{}

Two notes are in order here. First, Eqs.~\eqref{eq:Space_Interface_Scattered_Waves} explicitly shows how STESEM provides the \emph{complete and general scattering information}---the scattering amplitudes and the phase transformations for an arbitrary incident waveform---\emph{in a single step}, contrarily to the conventional approach that considers a specific waveform and determines the scattering phase in a first step and the scattering magnitude in a second step~\cite{Kong2008_Wave_Theory_BOOK,Jackson2012_Electro_BOOK}. Second, the expressions in Eqs.~\eqref{eq:Space_Interface_Scattered_Waves} may a priori look overly complicated for the problem considered. This is the case because of their total generality and exact phase reference, which will be more essential in the next problems. In the case of a simple plane wave incident on an interface located at $z_0=0$, we would have  here $\psi_{i}^{\pm}{\left[\tau_{i}^{\pm}\right]}=\exp\left(i\omega\tau_{i}^{\pm}\right)$ with $\tau_{i}^{\pm}=n_{i}z/c\mp t$ [Eq.~\eqref{eq:Traveling_Wave_Coordinates}], leading to the more familiar forms 
\begin{subequations}\label{eq:Space_Interface_Plane_Wave_Example}
    \begin{align}
        \psi_{1}^{-}{\left[n_{1}\frac{z}{c}+t\right]} &= \frac{\eta_{2}-\eta_{1}}{\eta_{2}+\eta_{1}}\exp\left(-i\omega\left(n_{1}\frac{z}{c}+t\right)\right) \,, \\
        \psi_{2}^{+}{\left[n_{2}\frac{z}{c}-t\right]} &= \frac{2\eta_{2}}{\eta_{2}+\eta_{1}}\exp\left(i\omega\left(n_{2}\frac{z}{c}-t\right)\right) \,,
    \end{align}
\end{subequations}
whose real parts provide the usual physical time-harmonic fields with the wavenumber $k_i=n_i\omega/c$.

\pati{Physics}{}

Figure~\ref{fig:Examples_Space-Time_Structures}a illustrates the scattering produced by a stationary space interface. Since the interface is stationary, the system possesses time-translation symmetry and the temporal frequency is therefore conserved across the interface, in agreement with the temporal Fourier spectra shown at the bottom of the figure. The reflected and transmitted amplitudes are determined solely by the impedance mismatch between the two media through the familiar Fresnel coefficients appearing in Eqs.~\eqref{eq:Space_Interface_Scattered_Waves}. In addition to these amplitude coefficients, Eqs.~\eqref{eq:Space_Interface_Scattered_Waves} contain constant offsets in the traveling-wave arguments. These offsets do not represent additional phase shifts introduced by the scattering process itself, but rather account for the propagation delays associated with the finite interface position for the interface placed at $z=z_0$. For the reflected wave, the offset~$2n_{1}z_{0}/c$ corresponds to the additional round-trip propagation delay: the incident pulse first propagates from the origin to the interface and then returns toward the observation point after reflection, resulting in an additional propagation distance of~$2z_{0}$. For the transmitted wave, propagation occurs successively in the two media. The incident pulse propagates with velocity~$c/n_{1}$ before reaching the interface and with velocity~$c/n_{2}$ after transmission. Consequently, the offset~$\left(n_{1}-n_{2}\right)z_{0}/c$ accounts for the difference between these two propagation delays. Although these phase offsets may appear unimportant in the present example, since one may simply set $z_{0} = 0$ [Eq.~\eqref{eq:Space_Interface_Plane_Wave_Example}], they become essential when considering multiple interfaces, such as the space-time wedge (Fig.~\ref{fig:Canonical_Space-Time_Structures}e), where the relative initial interface positions determine the correct phase relationships between successive scattering events.

\subsection{Time Interface}

\pati{Method}{}

The second canonical configuration is a purely temporal interface, in which the material properties change instantaneously and uniformly throughout space (Fig.~\ref{fig:Canonical_Space-Time_Structures}b). In contrast to the space interface, the scattering event occurs at a fixed time rather than at a fixed position. The incident wave generates two waves after the temporal transition: $\psi_{2}^{-}$  (later-backward) and $\psi_{2}^{+}$ (later-forward). Since the transition occurs instantaneously at every spatial position, the interface trajectory is described by~$t=t_{0}$, corresponding to the limiting case of an infinitely fast interface,~$v_{\text{m}}\rightarrow\infty$. All points in space therefore experience the material transition at the same instant, so the scattering event is uniquely identified by its spatial position,~$z^{\star}$, rather than by its scattering time. In this limit, the moving boundary conditions [Eqs.~\eqref{eq:General_Boundary_Conditions}] reduce to the continuity of the electric displacement and magnetic flux density fields at the temporal interface. Solving the boundary conditions at the interface time and subsequently extending the interface solution along the scattered-wave trajectories yields the complete space-time solution
\begin{subequations}\label{eq:Time_Interface_Scattered_Waves}
    \begin{align}
        \psi_{2}^{-}{\left[\tau_{2}^{-}\right]} &= \frac{n_{1}}{n_{2}}\frac{\eta_{2}-\eta_{1}}{2\eta_{1}}\psi_{\text{i}}\underbrace{\left[\frac{n_{1}}{n_{2}}\tau_{2}^{-} - \left(1+\frac{n_{1}}{n_{2}}\right)t_{0}\right]}_{\phi_{2}^{-}} \,, \\
        \psi_{2}^{+}{\left[\tau_{2}^{+}\right]} &= \frac{n_{1}}{n_{2}}\frac{\eta_{2}+\eta_{1}}{2\eta_{1}}\psi_{\text{i}}\underbrace{\left[\frac{n_{1}}{n_{2}}\tau_{2}^{+} - \left(1-\frac{n_{1}}{n_{2}}\right)t_{0}\right]}_{\phi_{2}^{+}} \,.
    \end{align}
\end{subequations}

\pati{Physics}{}

Figure~\ref{fig:Examples_Space-Time_Structures}b illustrates the temporal scattering solutions described by Eqs.~\eqref{eq:Time_Interface_Scattered_Waves}. Unlike a stationary space interface, a temporal interface breaks time-translation symmetry and therefore modifies the frequency content of the scattered waves. The factor~$n_{1}/n_{2}$ multiplying the waveform argument in both scattered waves gives rise to the temporal frequency scaling. This follows from differentiating the waveform phase with respect to time:~$\omega_{2}^{\pm}/\omega_{\text{i}} = -\partial \phi_{2}^{\pm} /\partial t=\pm n_{1}/n_{2}$, showing that both the later-forward and later-backward waves experience the same frequency scaling. Besides the amplitude change introduced by the impedance discontinuity, this scaling also stretches or compresses the waveform in time. Finally, the constant offsets in the waveform arguments account for the fact that the material transition occurs at~$t=t_{0}$. Their different forms reflect whether the scattered wave continues propagating in its original direction or reverses its direction following the temporal transition.

\subsection{Space-Time Corner}
\pati{Method}{}

The space-time corner consists of the combination of a space interface and a time interface, as illustrated in Fig.~\ref{fig:Canonical_Space-Time_Structures}c. Unlike the previously discussed canonical structures, the scattering process at a space-time corner cannot be described by a single scattering event. Instead, the incident waveform interacts with two distinct discontinuity boundaries, resulting in a sequence of coupled scattering processes. The first scattering event occurs when the incident waveform reaches the tip of the corner. Due to the simultaneous presence of spatial and temporal discontinuities, the incident field separates into multiple wave components. The spatial discontinuity generates conventional reflected and transmitted waves, whereas the temporal discontinuity generates the later-backward and later-forward waves associated with temporal scattering. The fields generated by the spatial interface are directly obtained from the space-interface solutions in Eqs.~\eqref{eq:Space_Interface_Scattered_Waves}, while the fields generated by the temporal interface follow from the time-interface solutions in Eqs.~\eqref{eq:Time_Interface_Scattered_Waves}. The later-backward wave from the time interface propagates towards the spatial interface and undergoes, upon reaching this interface, a second scattering event, producing an additional reflected and transmitted contribution. This secondary scattering process is governed by the space-interface solutions in Eqs.~\eqref{eq:Space_Interface_Scattered_Waves}, with the medium indices reversed since the wave now propagates from the second medium into the first medium. Overall, the complete scattered solution is obtained as the superposition of all generated wave components. This decomposition illustrates the modular nature of STESEM, where complex space-time scattering problems can be systematically constructed from elementary scattering events with known solutions.

\pati{Physics}{}

The space-time corner combines the two fundamental symmetry-breaking mechanisms introduced previously: spatial and temporal discontinuities. The wave component interacting with the spatial interface undergoes conventional Fresnel scattering. Since the interface is stationary, time-translation symmetry is preserved and the frequency remains unchanged. Conversely, the component interacting with the temporal interface experiences a temporal discontinuity. Because the system remains spatially invariant, spatial-translation symmetry is preserved and the momentum remains unchanged, while the frequency is modified. The resulting frequency conversion follows the temporal scaling~($n_{1}/n_{2}$), as described by Eqs.~\eqref{eq:Time_Interface_Scattered_Waves}. Therefore, the scattered spectrum contains contributions at two distinct frequencies: the original frequency associated with the spatial scattering pathway and the shifted frequency generated by the temporal scattering pathway. This spectral splitting is visible in Fig.~\ref{fig:Examples_Space-Time_Structures}c, where  the reflected and transmitted fields exhibit distinct spectral peaks.

\subsection{Constant-Velocity Traveling Interface} 
\pati{Method}{}

For a constant-velocity traveling interface in the subluminal regime, the scattered waves consist of~$\psi_{1}^{-}$ (reflection) and~$\psi_{2}^{+}$ (transmission), see Fig.~\ref{fig:Canonical_Space-Time_Structures}d. The interface trajectory is described by~$z = v_{\text{m}}t + z_{0}$. Unlike the previous canonical structures, the scattering position and scattering time are no longer independent variables, since they are constrained by the trajectory of the moving interface. Consequently, the scattering event associated with an arbitrary observation point must be determined by finding the intersection between the corresponding wave trajectory [Eq.~\eqref{eq:Space_Interface_Trajectory_Pulse}] and the interface trajectory. Solving this intersection provides the scattering time and position associated with each observation point. Substituting these scattering coordinates into the scattering coefficients of the moving interface yields the complete space-time solution
\begin{subequations}\label{eq:Constant_Interface_Scattered_Waves}
    \begin{align}
        &\psi_{1}^{-}{\left[\tau_{1}^{-}\right]} = \frac{\eta_{2}-\eta_{1}}{\eta_{2}+\eta_{1}}\frac{1-n_{1}v_{\text{m}}/c}{1+n_{1}v_{\text{m}}/c} \nonumber \\
        &\hspace{0cm} \times\psi_{\text{i}}{\left[-\frac{1-n_{1}v_{\text{m}}/c}{1+n_{1}v_{\text{m}}/c}\tau_{1}^{-} + n_{1}\left(1+\frac{1-n_{1}v_{\text{m}}/c}{1+n_{1}v_{\text{m}}/c}\right)\frac{z_{0}}{c}\right]}\,, \\
        &\psi_{2}^{+}{\left[\tau_{2}^{+}\right]} = \frac{2\eta_{2}}{\eta_{2}+\eta_{1}}\frac{1-n_{1}v_{\text{m}}/c}{1-n_{2}v_{\text{m}}/c} \nonumber \\
        &\hspace{0cm} \times\psi_{\text{i}}{\left[\frac{1-n_{1}v_{\text{m}}/c}{1-n_{2}v_{\text{m}}/c}\tau_{2}^{+} + \left(n_{1}-n_{2}\frac{1-n_{1}v_{\text{m}}/c}{1-n_{2}v_{\text{m}}/c}\right)\frac{z_{0}}{c}\right]}\,.
    \end{align}
\end{subequations}

\pati{Physics}{}

Figure~\ref{fig:Examples_Space-Time_Structures}d illustrates the scattering produced by a constant moving interface. The scattering amplitudes contain two contributions: the conventional impedance-mismatch coefficients associated with a stationary interface ($v_{\text{m}} = 0$) and additional Doppler factors resulting from the interface motion. The impedance mismatch determines the relative amplitudes of the reflected and transmitted waves, whereas the Doppler factors modify the temporal evolution of the scattered waveforms through stretching or compression. Specifically, the factors multiplying the traveling-wave variables represent the Doppler scaling, viz. $\omega_{i}^{\pm}/\omega_{\text{i}} = \left(n_{1}v_{\text{m}}/c-1\right)/\left(n_{i}v_{\text{m}}/c\mp 1\right)$, which changes the observed frequency and consequently leads to spectral compression or expansion. Finally, the constant offsets appearing in Eqs.~\eqref{eq:Constant_Interface_Scattered_Waves} correspond to phase shifts associated with the initial interface position. These terms account for the propagation delays accumulated before and after the scattering event and ensure the correct phase relationship between the incident and scattered wave components.

\subsection{Space-Time Wedge}
\pati{Method}{
}

A space-time wedge consists of two traveling interfaces enclosing some space-time region, as shown in Fig.~\ref{fig:Canonical_Space-Time_Structures}e. The interfaces are parameterized as~$z=v_{\text{m},i}t + z_{i}$, where~$v_{\text{m},i}$ denotes the modulation velocity and~$z_{i}$ the initial position of interface~$i = 1, 2$. In contrast to the previous canonical structures, a wave interacting with a wedge experiences multiple successive scattering events. After the first interaction with one interface, the generated wave components propagate towards the second interface, where they undergo additional scattering processes. These subsequent interactions are described by the constant-velocity traveling-interface solutions of Eqs.~\eqref{eq:Constant_Interface_Scattered_Waves}, with the resulting scattered waves acting as incident fields for further interactions. If the wedge geometry permits internal reflections, higher-order scattering pathways are generated through repeated interactions between the two interfaces. Within STESEM, all interface interactions are treated simultaneously, but may also be understood as elementary scattering events whose solution are already known. The total scattered field is then given by coherently superposing all possible scattering pathways.

\pati{Physics}{}

The space-time wedge illustrates one of the principal advantages of STESEM: complex space-time structures can be systematically analyzed by combining elementary scattering events. Each interaction with a moving interface introduces a Doppler transformation determined by the refractive index contrast and modulation velocity [Eqs.~\eqref{eq:Constant_Interface_Scattered_Waves}]. Therefore, a wave propagating through the wedge does not experience a single frequency conversion, but rather a sequence of frequency transformations accumulated along its scattering pathway. Different propagation pathways correspond to different sequences of Doppler transformations, resulting in multiple frequency components in the scattered spectrum associated with the different possible interaction histories. Internal reflections within the wedge introduce additional higher-order pathways, further enriching the spectral response. This behavior is visible in Fig.~\ref{fig:Examples_Space-Time_Structures}e, where multiple spectral peaks appear in the reflected spectrum, each corresponding to a distinct scattering event.

\subsection{Accelerated Interface}
\pati{Method}{}

We finally consider the most general canonical structure discussed in this tutorial: an arbitrary accelerated subluminal interface, whose velocity varies continuously with time, as illustrated in Fig.~\ref{fig:Canonical_Space-Time_Structures}f. The interface trajectory is described by an arbitrary function~$z=z{\left[t\right]}$ and the scattered waves are given by~$\psi_{1}^{-}$ and~$\psi_{2}^{+}$. Unlike the constant-velocity case, the scattering time can no longer be obtained explicitly by intersecting the wave trajectory with the interface trajectory. Instead, the scattering event is implicitly determined by the nonlinear interface motion. To establish this relation, we introduce the auxiliary functions
\begin{equation}\label{eq:Accelerated_Interface_Auxiliary_Functions}
    f_{i}^{\pm}{\left[t^{\star}\right]} = n_{i}\frac{z{\left[t^{\star}\right]}}{c}\mp t^{\star}\,,
\end{equation}
which represent the traveling-wave coordinates [Eq.~\eqref{eq:Traveling_Wave_Coordinates}] evaluated at the scattering event~$\left(z^{\star},ct^{\star}\right)$. Thus,~$f_{i}^{\pm}$ maps each scattering time to the corresponding traveling-wave coordinate. The inverse mapping,~$\left(f_{i}^{\pm}\right)^{-1}$, then recovers the scattering time associated with each observation coordinate, as the traveling wave is invariant along the scattered wave trajectories. The complete scattered fields are obtained as
\begin{subequations}\label{eq:Accelerated_Interface_Scattered_Waves}
    \begin{align}
        \psi_{1}^{-}{\left[\tau_{1}^{-}\right]} &= \frac{\eta_{2}-\eta_{1}}{\eta_{2}+\eta_{1}}\frac{1-n_{1}v_{\text{m}}{\left[\left(f_{1}^{-}\right)^{-1}{\left[\tau_{1}^{-}\right]}\right]}/c}{1+n_{1}v_{\text{m}}{\left[\left(f_{1}^{-}\right)^{-1}{\left[\tau_{1}^{-}\right]}\right]}/c} \nonumber \\
        &\hspace{2.3cm} \times\psi_{\text{i}}{\left[f_{1}^{+}{\left[\left(f_{1}^{-}\right)^{-1}{\left[\tau_{1}^{-}\right]}\right]}\right]}\,, \\
        \psi_{2}^{+}{\left[\tau_{2}^{+}\right]} &= \frac{2\eta_{2}}{\eta_{2}+\eta_{1}}\frac{1-n_{1}v_{\text{m}}{\left[\left(f_{2}^{+}\right)^{-1}{\left[\tau_{2}^{+}\right]}\right]}/c}{1-n_{2}v_{\text{m}}{\left[\left(f_{2}^{+}\right)^{-1}{\left[\tau_{2}^{+}\right]}\right]}/c} \nonumber \\
        &\hspace{2.3cm} \times\psi_{\text{i}}{\left[f_{1}^{+}{\left[\left(f_{2}^{+}\right)^{-1}{\left[\tau_{2}^{+}\right]}\right]}\right]}\,.
    \end{align}
\end{subequations}

\pati{Physics}{}

As the most general canonical structure considered in this tutorial, the accelerated interface demonstrates the general capability of STESEM, as shown in Fig.~\ref{fig:Examples_Space-Time_Structures}f. The physical ingredients in Eqs.~\eqref{eq:Accelerated_Interface_Scattered_Waves} remain identical to those encountered for the constant-velocity interface [Eqs.~\eqref{eq:Constant_Interface_Scattered_Waves}]: the impedance mismatch determines part of the scattering amplitudes, whereas Doppler transformations govern the temporal evolution of the scattered waveforms. The essential difference is that the interface velocity is no longer constant but instead depends on the scattering time associated with each observation point. Consequently, different portions of the waveform interact with different instantaneous interface velocities and experience different Doppler shifts. Specifically, the local Doppler frequency scalings are given by $\omega_{i}^{\pm}/\omega_{\text{i}} = \left(n_{1}v_{\text{m}}{\left[t^{\star}\right]}/c - 1\right)/\left(n_{i}v_{\text{m}}{\left[t^{\star}\right]}/c \mp 1\right)$, where~$t^{\star} = \left(f_{i}^{\pm}\right)^{-1}{\left[\tau_{i}^{\pm}\right]}$ is the scattering time associated with the observation point. Since~$v_{\text{m}}{\left[t^{\star}\right]}$ varies continuously across the waveform, the Doppler scaling likewise varies continuously, producing a continuously varying instantaneous frequency. This phenomenon manifests itself as frequency chirping, in which different temporal portions of the scattered waveform undergo different frequency shifts according to the instantaneous interface velocity at their respective scattering events. The inverse auxiliary mappings in Eqs.~\eqref{eq:Accelerated_Interface_Scattered_Waves} determine these scattering events and automatically provide the corresponding local Doppler transformations. Therefore, even for arbitrary nonlinear interface trajectories, STESEM retains the same fundamental interpretation established throughout this tutorial: the scattering event is first determined, after which the resulting waveform is extended throughout space-time along its invariant traveling-wave coordinate.

\pati{Scheme Numerical Solving Solutions}{}

Equations~\eqref{eq:Accelerated_Interface_Scattered_Waves} may appear complicated because they involve inverse auxiliary mappings, which generally do not admit closed-form expressions for arbitrary interface trajectories. Although particular cases, such as interfaces undergoing uniform proper acceleration, can be treated analytically, the resulting expressions rapidly become cumbersome. In practice, \emph{numerical evaluation} is therefore both more convenient and considerably simpler. The procedure is straightforward: 
\begin{enumerate}
    \item[(i)] construct the spatial~($z$) and temporal~($t$) vector arrays that span the desired space-time;
    \item[(ii)] compute the corresponding traveling-wave vector arrays~$\tau_{1}^{-}$ and~$\tau_{2}^{+}$ using Eq.~\eqref{eq:Traveling_Wave_Coordinates};
    \item[(iii)] define the interface function~$z{\left[t\right]}$;
    \item[(iv)] construct the corresponding auxiliary functions~$f_{i}^{\pm}{\left[t^{\star}\right]}$ in Eqs.~\eqref{eq:Accelerated_Interface_Auxiliary_Functions} that includes $z{\left[t\right]}$ in~(iii);
    \item[(v)] numerically invert these functions using a standard inversion routine over the~$\tau_{i}^\pm$-arrays in~(ii) to determine the corresponding scattering times,~$t^{\star}=\left(f_{i}^{\pm}\right)^{-1}\left[\tau_{i}^{\pm}\right]$;
    \item[(vi)] calculate the interface velocity function~$v_{\text{m}}{\left[t\right]}$ by (numerically) differentiating~$z{\left[t\right]}$ in~(iii);
    \item[(vii)] insert~$t^{\star}$ from~(v) into~$f_{i}^{\pm}{\left[t^\star\right]}$ in~(iv) and $v_{\text{m}}{\left[t\right]}$ in (vi) into Eqs.~\eqref{eq:Accelerated_Interface_Scattered_Waves}.
\end{enumerate}

\subsection{Other, Arbitrary Interfaces}
Although this tutorial has focused on the six canonical structures shown in Fig.~\ref{fig:Canonical_Space-Time_Structures}, STESEM is fully general and applies to arbitrary space-time interfaces, including complex trajectories such as the one illustrated in Fig.~\ref{fig:Problem_statement_and_STEM}d.

%%%%%%%%%%%%%%%%%%%%%%%%%%%%%%%%%%%%%%%%%%%%%%%%%%%%%%%%%%%%%%%%%%%%%%%%%%%%%%%%%%%%%%%%%%%%%%%%%%%%
\section{Conclusions}\label{sec:Conclusions}

\pati{Summary}{}

In this tutorial, we have introduced STESEM as a universal analytical framework for solving electromagnetic scattering problems at arbitrary space-time interfaces directly in the laboratory frame. Unlike conventional approaches, which typically require problem-specific analytical derivations or transformations to dedicated reference frames, STESEM provides a unified methodology applicable to a broad range of canonical space-time configurations. We have demonstrated the framework by deriving the scattering response of canonical space-time structures and have shown how more complex configurations can be systematically decomposed into elementary scattering events. Beyond the analytical derivations, the presentation has highlighted the physical mechanisms underlying wave scattering at moving interfaces, including momentum transitions, frequency conversion, multiple Doppler shifting and frequency chirping. We hope that resulting framework provides researchers with both a practical analytical tool and an intuitive physical framework for analyzing and designing wave phenomena in space-time metamaterials.

\pati{Extensions}{}

Beyond the systems considered in this tutorial, STESEM can be extended towards increasingly general classes of space-time electromagnetic systems. These extensions include inverse space-time design problems, such as arbitrary pulse shaping~\cite{Bahrami2025_Pulse_Shap_ASTEM_PUB} and generalized frequency chirping~\cite{DeKinder2026_Scat_Chirp_ASTEM_PUB}, which generally require accelerated interfaces. The inclusion of material dispersion~\cite{DeKinder2026_Scat_ST_Int_Disp_USTEM} introduces additional propagating modes and enables the study of novel phenomena such as space-time focusing~\cite{Ostrovskii1975_Lens_ASTEM_PUB,Dexelle2026_ST_Lensing}. Furthermore, extending STESEM to bianisotropic media enables the analysis of more general classes of space-time interfaces and metamaterials with coupled electric and magnetic responses. The framework can also be extended and applied to multiple interacting interfaces, including periodic, homogenized and aperiodic structures, ultimately enabling the analysis and design of arbitrary accelerated space-time crystals. Finally, the generalization towards higher-dimensional space-time structures, including 2+1D and 3+1D configurations, will enable the treatment of multidimensional wave manipulation and more realistic electromagnetic systems.

\appendix

\section{Scattering at Canonical Space-Time Structures}\label{sec:appendix:Scattering_at_Canonical_Space-Time_Structures}
    \pati{Purpose Appendix}{
    }

    This appendix provides detailed space-time event scattering and extension method (STESEM) mathematical derivations of the scattered electromagnetic fields for all the canonical space-time structures in Fig.~\ref{fig:Canonical_Space-Time_Structures}. STESEM decomposes the scattering problem into two complementary steps. In the first step, the waves scattered at the interface are determined by formulating and solving the moving boundary conditions. In the second step, the resulting interface solutions are extended to arbitrary space-time points by tracing each space-time point back to its corresponding scattering event, leveraging the invariance of the traveling-wave coordinate along the scattered wave trajectories.

    \subsection{Problem Statement}\label{subsec:appendix:Problem_Statement}
        \pati{General Assumptions}{
        }

        We consider a one-dimensional electromagnetic system consisting of a moving abrupt interface separating two isotropic, linear and nondispersive media. The two media are characterized by refractive indices~$n_{i}$ and impedances~$\eta_{i}$, where $i=1,2$ denotes the medium. The interface moves along the $z$-direction and is described by the trajectory~$z{\left[t\right]}$, with instantaneous velocity~$v_{\text{m}}{\left[t\right]}=\dd{z}/\dd{t}$. The electromagnetic fields are assumed to depend solely on~$z$ and~$t$. A transverse electromagnetic (TEM) plane wave is assumed, where the electric field is polarized along the $x$-direction and the magnetic field along the $y$-direction. General forward-and backward-propagating electromagnetic waveforms in each medium are expressed as
        \begin{subequations}\label{eq:appendix:General_Traveling_Wave_Functions}
            \begin{align}
                E_{i}^{\pm} &= \psi_{i}^{\pm}{\left[\tau_{i}^{\pm}\right]}\,, &H_{i}^{\pm} &= \pm \frac{1}{\eta_{i}}\psi_{i}^{\pm}{\left[\tau_{i}^{\pm}\right]}\,, \\
                D_{i}^{\pm} &= \frac{n_{i}}{c\eta_{i}}\psi_{i}^{\pm}{\left[\tau_{i}^{\pm}\right]}\,, &B_{i}^{\pm} &= \pm \frac{n_{i}}{c}\psi_{i}^{\pm}{\left[\tau_{i}^{\pm}\right]}\,.
            \end{align}
        \end{subequations}
        In these relations,~$\psi_{i}^{\pm}$ denotes an arbitrary waveform and the square brackets include its argument, which takes the traveling-wave form
        \begin{equation}\label{eq:appendix:Traveling_Wave_Variables}
            \tau_{i}^{\pm} = n_{i}\frac{z}{c}\mp t\,,
        \end{equation}
        where the symbol~$\pm$ denotes forward~($+$) or backward~($-$) propagating waves, respectively, while the constant~$c$ represents the speed of light in vacuum.
    
    \subsection{General Boundary Conditions}
        \pati{General Boundary Conditions}{
        }

        The electromagnetic moving boundary conditions at the interface,~$z{\left[t\right]}$, are given by~\cite{Pauli1981_Relativity_BOOK,Caloz2019b_ST_Metamaterials_USTEM_PUB}
        \begin{subequations}\label{eq:appendix:General_Boundary_Conditions}
            \begin{align}
                \left.E_{1}^{\pm}-v_{\text{m}}{\left[t^{\star}\right]}B_{1}^{\pm}\right|_{z=z{\left[t^{\star}\right]}} &= \left.E_{2}^{\pm}-v_{\text{m}}{\left[t^{\star}\right]}B_{2}^{\pm}\right|_{z=z{\left[t^{\star}\right]}} \,, \\
                \left.H_{1}^{\pm} - v_{\text{m}}{\left[t^{\star}\right]}D_{1}^{\pm}\right|_{z=z{\left[t^{\star}\right]}} &= \left. H_{2}^{\pm} - v_{\text{m}}{\left[t^{\star}\right]}D_{2}^{\pm}\right|_{z=z{\left[t^{\star}\right]}}\,,
            \end{align}
        \end{subequations}
        where $t^{\star}$ represents the scattering time at the interface (Fig.~\ref{fig:Problem_statement_and_STEM}d). Substituting Eqs.~\eqref{eq:appendix:General_Traveling_Wave_Functions} into Eqs.~\eqref{eq:appendix:General_Boundary_Conditions} with traveling-wave coordinates [Eq.~\eqref{eq:appendix:Traveling_Wave_Variables}] evaluated along the interface trajectory~$z{\left[t^{\star}\right]}$ gives the general moving boundary conditions
        \begin{widetext}
            \begin{subequations}\label{eq:appendix:Boundary_Conditions_Psi}
                \begin{align}
                    \left(1 \mp n_{1}\frac{v_{\text{m}}{\left[t^{\star}\right]}}{c}\right)\psi_{1}^{\pm}{\left[n_{1}\frac{z{\left[t^{\star}\right]}}{c}\mp t^{\star}\right]} &= \left(1 \mp n_{2}\frac{v_{\text{m}}{\left[t^{\star}\right]}}{c}\right)\psi_{2}^{\pm}{\left[n_{2}\frac{z{\left[t^{\star}\right]}}{c}\mp t^{\star}\right]} \,, \\
                    \pm\frac{1}{\eta_{1}}\left(1 \mp n_{1}\frac{v_{\text{m}}{\left[t^{\star}\right]}}{c}\right)\psi_{1}^{\pm}{\left[n_{1}\frac{z{\left[t^{\star}\right]}}{c}\mp t^{\star}\right]} &= \pm\frac{1}{\eta_{2}}\left(1 \mp n_{2}\frac{v_{\text{m}}{\left[t^{\star}\right]}}{c}\right)\psi_{2}^{\pm}{\left[n_{2}\frac{z{\left[t^{\star}\right]}}{c}\mp t^{\star}\right]}\,.
                \end{align}
            \end{subequations}
        \end{widetext}
        For each canonical scattering configuration, the total field on either side of the interface is obtained by summing the corresponding non-zero wave components, as will become apparent shortly. Equations~\eqref{eq:appendix:Boundary_Conditions_Psi} may appear a priori abstract, but they are very powerful as they constitute the general moving boundary conditions for an arbitrary interface trajectory. They provide the field solutions at the interface once the incident waveform and interface trajectory have been specified. The following sections analyze the scattering of an incident wave propagating from the first medium onto the second medium for the canonical space-time structures shown in Fig.~\ref{fig:Canonical_Space-Time_Structures}.

    \subsection{Space (Stationary) Interface}\label{subsec:appendix:Space_Interface}
        \pati{Scattered Waves}{
        }

        For a space interface, the scattered waves are~$\psi_{1}^{-}$ (reflection) and~$\psi_{2}^{+}$ (transmission), see Fig.~\ref{fig:Canonical_Space-Time_Structures}a. Since the interface does not move, its velocity is zero ($v_{\text{m}} = 0$) and hence, the interface trajectory is simply described by $z=z_{0}$. Consequently, the boundary conditions reduce to the continuity of $E_{i}^{\pm}$ and $H_{i}^{\pm}$ [Eq.~\eqref{eq:appendix:General_Boundary_Conditions}]. Setting $v_{\text{m}}=0$ and using $z{\left[t^{\star}\right]} = z_{0}$ in Eq.~\eqref{eq:appendix:Boundary_Conditions_Psi} results in 
        \begin{subequations}\label{eq:appendix:Space_Interface_Boundary_Conditions}
            \begin{align}
                \underbrace{\psi_{\text{i}}{\left[n_{1}\frac{z_{0}}{c} - t^{\star}\right]}}_{\text{incident}} + \underbrace{\psi_{1}^{-}{\left[n_{1}\frac{z_{0}}{c} + t^{\star}\right]}}_{\text{reflection}} &= \underbrace{\psi_{2}^{+}{\left[n_{2}\frac{z_{0}}{c} - t^{\star}\right]}}_{\text{transmission}}\,, \\
                \frac{1}{\eta_{1}}\psi_{\text{i}}{\left[n_{1}\frac{z_{0}}{c} - t^{\star}\right]} - \frac{1}{\eta_{1}}\psi_{1}^{-}{\left[n_{1}\frac{z_{0}}{c} + t^{\star}\right]} &= \frac{1}{\eta_{2}}\psi_{2}^{+}{\left[n_{2}\frac{z_{0}}{c} - t^{\star}\right]}\,,
            \end{align}
        \end{subequations}
        where~$\psi_{\text{i}}$ is the incident wave traveling forward in the first medium. Solving these equations for $\psi_{1}^{-}$ and $\psi_{2}^{+}$ yields then
        \begin{subequations}\label{eq:appendix:Space_Interface_Solution_at_Interface}
            \begin{align}
                \psi_{1}^{-}{\left[n_{1}\frac{z_{0}}{c} + t^{\star}\right]} &= \frac{\eta_{2}-\eta_{1}}{\eta_{2}+\eta_{1}}\psi_{\text{i}}{\left[n_{1}\frac{z_{0}}{c} - t^{\star}\right]}\,, \\
                \psi_{2}^{+}{\left[n_{2}\frac{z_{0}}{c} - t^{\star}\right]} &= \frac{2\eta_{2}}{\eta_{2}+\eta_{1}}\psi_{\text{i}}{\left[n_{1}\frac{z_{0}}{c} - t^{\star}\right]}\,.
            \end{align}
        \end{subequations}
        Equations~\eqref{eq:appendix:Space_Interface_Solution_at_Interface} determine the scattered fields only at the interface. To obtain the complete space-time solution, the interface solutions must be extended away from the interface. Once scattered, each scattered wave propagates independently within its corresponding homogeneous medium. Therefore, the traveling-wave coordinate remains constant along the wave trajectory, and the waveform value remains unchanged. Consider an arbitrary observation point, $\left(z_{i}^{\pm},ct_{i}^{\pm}\right)$ (Fig.~\ref{fig:Problem_statement_and_STEM}d). The corresponding scattering event is obtained by tracing the wave trajectory backward until it intersects the interface trajectory. The trajectory of a wave passing through this point is (Fig.~\ref{fig:Problem_statement_and_STEM}d)
        \begin{equation}\label{eq:appendix:Space_Interface_Trajectory_Pulse}
            \begin{split}
                z &= \pm \frac{c}{n_{i}}\left(t-t_{i}^{\pm}\right)+z_{i}^{\pm} \\
                &= \pm \frac{c}{n_{i}}t + \frac{c}{n_{i}}\underbrace{\left(n_{i}\frac{z_{i}^{\pm}}{c}\mp t_{i}^{\pm}\right)}_{\tau_{i}^{\pm}}\,.
            \end{split}
        \end{equation}
        The term in parentheses is precisely the traveling-wave coordinate of the observation point~$\tau_{i}^{\pm}$ [Eq.~\eqref{eq:appendix:Traveling_Wave_Variables}], which remains constant along the wave trajectory. Equating~\eqref{eq:appendix:Space_Interface_Trajectory_Pulse} with the (stationary) interface position,~$z_{0}$, gives the scattering time associated with the observation point,
        \begin{equation}\label{eq:appendix:Space_Interface_Coordinate_Transformation}
            t^{\star} = \pm\left(n_{i}\frac{z_{0}}{c}-\tau_{i}^{\pm}\right)\,.
        \end{equation}
        Equation~\eqref{eq:appendix:Space_Interface_Coordinate_Transformation} establishes a one-to-one correspondence between every observation point and its associated scattering event, and represents the crux of the new method. Substituting this equation into Eqs.~\eqref{eq:appendix:Space_Interface_Solution_at_Interface} provides the complete scattered fields
        \begin{subequations}\label{eq:appendix:Space_Interface_Scattered_Waves}
            \begin{align}
                \psi_{1}^{-}{\left[\tau_{1}^{-}\right]} &= \frac{\eta_{2}-\eta_{1}}{\eta_{2}+\eta_{1}}\psi_{\text{i}}{\left[-\tau_{1}^{-}+2n_{1}\frac{z_{0}}{c}\right]}\,, \\
                \psi_{2}^{+}{\left[\tau_{2}^{+}\right]} &= \frac{2\eta_{2}}{\eta_{2}+\eta_{1}}\psi_{\text{i}}{\left[\tau_{2}^{+} + \left(n_{1}-n_{2}\right)\frac{z_{0}}{c}\right]}\,.
            \end{align}
        \end{subequations}

    \subsection{Time (Instantaneous) Interface}\label{subsec:appendix:Time_Interface}

        \pati{Scattered Waves}{
        }

        At a time interface, the scattered waves are~$\psi_{2}^{-}$ (later-backward) and~$\psi_{2}^{+}$ (later-forward), as shown in Fig.~\ref{fig:Canonical_Space-Time_Structures}b. The interface occurs instantaneously throughout space and is therefore described by $t = t_{0}$, corresponding to the limiting case $v_{\text{m}} \rightarrow \infty$. Consequently, the scattering event is identified by its spatial position $z^{\star}$, rather than its time coordinate. In this case, the boundary conditions reduce to the continuity of $D_{i}^{\pm}$ and $B_{i}^{\pm}$ [Eq.~\eqref{eq:appendix:General_Boundary_Conditions}]. Application of Eq.~\eqref{eq:appendix:Boundary_Conditions_Psi} result in
        \begin{subequations}\label{eq:appendix:Time_Interface_Boundary_Conditions}
            \begin{align}
                n_{1}\psi_{\text{i}}{\left[n_{1}\frac{z^{\star}}{c} - t_{0}\right]}  &= -n_{2}\psi_{2}^{-}{\left[n_{2}\frac{z^{\star}}{c} + t_{0}\right]} \nonumber \\
                &\hspace{1cm} + n_{2}\psi_{2}^{+}{\left[n_{2}\frac{z^{\star}}{c} - t_{0}\right]}\,, \\
                \frac{n_{1}}{\eta_{1}}\psi_{\text{i}}{\left[n_{1}\frac{z^{\star}}{c} - t_{0}\right]} &= \frac{n_{2}}{\eta_{2}}\psi_{2}^{-}{\left[n_{2}\frac{z^{\star}}{c} + t_{0}\right]} \nonumber \\
                &\hspace{1cm} + \frac{n_{2}}{\eta_{2}}\psi_{2}^{+}{\left[n_{2}\frac{z^{\star}}{c} - t_{0}\right]}\,.
            \end{align}
        \end{subequations}
        Solving Eqs.~\eqref{eq:appendix:Time_Interface_Boundary_Conditions} for $\psi_{2}^{\pm}$ yields
        \begin{subequations}\label{eq:appendix:Time_Interface_Solution_at_Interface}
            \begin{align}
                \psi_{2}^{-}{\left[n_{2}\frac{z^{\star}}{c} + t_{0}\right]} &= \frac{n_{1}}{n_{2}}\frac{\eta_{2}-\eta_{1}}{2\eta_{1}}\psi_{\text{i}}{\left[n_{1}\frac{z^{\star}}{c} - t_{0}\right]}\,, \\
                \psi_{2}^{+}{\left[n_{2}\frac{z^{\star}}{c} - t_{0}\right]} &= \frac{n_{1}}{n_{2}}\frac{\eta_{2}+\eta_{1}}{2\eta_{1}}\psi_{\text{i}}{\left[n_{1}\frac{z^{\star}}{c} - t_{0}\right]}\,.
            \end{align}
        \end{subequations}
        Again, these expressions describe only the fields at the (instantaneous) interface. The STESEM procedure is therefore applied to provide the complete space-time solution. The wave trajectory of the scattered waves can be written as (Fig.~\ref{fig:Problem_statement_and_STEM}d)
        \begin{equation}\label{eq:appendix:Time_Interface_Trajectory_Pulse}
            t = \pm n_{i}\frac{z}{c}\mp\tau_{i}^{\pm}\,.
        \end{equation}
        Intersecting Eq.~\eqref{eq:appendix:Time_Interface_Trajectory_Pulse} with the temporal interface $t=t_{0}$ and solving for the scattering position gives
        \begin{equation}\label{eq:appendix:Time_Interface_Coordinate_Transformation}
            z^{\star} = \pm\frac{c}{n_{i}}\left(t_{0}\pm\tau_{i}^{\pm}\right)\,.
        \end{equation}
        Substitution of Eq.~\eqref{eq:appendix:Time_Interface_Coordinate_Transformation} into Eqs.~\eqref{eq:appendix:Time_Interface_Solution_at_Interface} finally provides the scattered fields throughout space-time
        \begin{subequations}\label{eq:appendix:Time_Interface_Scattered_Waves}
            \begin{align}
                \psi_{2}^{-}{\left[\tau_{2}^{-}\right]} &= \frac{n_{1}}{n_{2}}\frac{\eta_{2}-\eta_{1}}{2\eta_{1}}\psi_{\text{i}}{\left[\frac{n_{1}}{n_{2}}\tau_{2}^{-} - \left(1+\frac{n_{1}}{n_{2}}\right)t_{0}\right]}\,, \\
                \psi_{2}^{+}{\left[\tau_{2}^{+}\right]} &= \frac{n_{1}}{n_{2}}\frac{\eta_{2}+\eta_{1}}{2\eta_{1}}\psi_{\text{i}}{\left[\frac{n_{1}}{n_{2}}\tau_{2}^{+} - \left(1-\frac{n_{1}}{n_{2}}\right)t_{0}\right]}\,.
            \end{align}
        \end{subequations}

    \subsection{Space-time Corner}\label{subsec:appendix:Space-Time_Corner}
        \pati{Explanation Two Scattering Events}{
        }

        The space-time corner combines the space interface (Sec.~\ref{subsec:appendix:Space_Interface}) and the time interface (Sec.~\ref{subsec:appendix:Time_Interface}), as illustrated in Fig.~\ref{fig:Canonical_Space-Time_Structures}c. Unlike the previous canonical structures, a space-time corner cannot be described by a single scattering event. Instead, the interaction proceeds through two successive scattering events. The first scattering event occurs when the incident waveform encounters the space-time corner: a portion of the incident waveform interacts with the time interface, whereas the remaining portion interacts with the space interface. Consequently, the first scattering event generates the conventional~$\psi_{2}^{-\text{T}}$ (later-backward) and~$\psi_{2}^{+\text{T}}$ (later-forward) waves associated with the temporal interface as well as the~$\psi_{1}^{-\text{S}}$ (reflection) and~$\psi_{2}^{+\text{S}}$ (transmission) waves associated with the stationary interface. The second scattering event originates from~$\psi_{2}^{-\text{T}}$ generated during the temporal interaction. This wave propagates toward the stationary interface and subsequently undergoes a second scattering process. This additional interaction produces~$\psi_{2}^{+\text{ST}}$ (reflection) and~$\psi_{1}^{-\text{ST}}$ (transmission). The complete scattered fields are therefore
       \begin{subequations}\label{eq:appendix:ST_Corner_Sum_Scattered_Waves}
            \begin{align}
                \psi_{1}^{-}{\left[\tau_{1}^{-}\right]} &= \psi_{1}^{-\text{S}}{\left[\tau_{1}^{-}\right]} + \psi_{1}^{-\text{ST}}{\left[\tau_{1}^{-}\right]} \,, \\
                \psi_{2}^{+}{\left[\tau_{2}^{+}\right]} &= \psi_{2}^{+\text{S}}{\left[\tau_{2}^{+}\right]} + \psi_{2}^{+\text{T}}{\left[\tau_{2}^{+}\right]} + \psi_{2}^{+\text{ST}}{\left[\tau_{2}^{+}\right]}\,,
            \end{align}
        \end{subequations}
        where the superscripts~S,~T, and~ST denote waves generated by the stationary interface, the temporal interface, and the second scattering event (time interface followed by space interface), respectively.

        \pati{First Scattering Event}{
        }
        
        To determine the contributions associated with the first scattering event, we decompose the incident waveform into two parts using the Heaviside step function, $\Theta{\left[\cdot\right]}$:
        \begin{equation}\label{eq:appendix:ST_Corner_Splitting_Incident_Wave}
            \psi_{\text{i}}{\left[\tau_{1}^{+}\right]} = \psi_{\text{i}}^{\text{S}}{\left[\tau_{1}^{+}\right]} + \psi_{\text{i}}^{\text{T}}{\left[\tau_{1}^{+}\right]}\,,
        \end{equation}
        where
        \begin{subequations}
            \begin{align}
                \psi_{\text{i}}^{\text{S}}{\left[\tau_{1}^{+}\right]} &= \psi_{\text{i}}{\left[\tau_{1}^{+}\right]}\Theta{\left[\tau_{0}-\tau_{1}^{+}\right]} \,, \label{eq:appendix:ST_Corner_Splitting_Incident_Wave_Space} \\
                \psi_{\text{i}}^{\text{T}}{\left[\tau_{1}^{+}\right]} &= \psi_{\text{i}}{\left[\tau_{1}^{+}\right]}\Theta{\left[\tau_{1}^{+}-\tau_{0}\right]} \,. \label{eq:appendix:ST_Corner_Splitting_Incident_Wave_Time} 
            \end{align}
        \end{subequations}
        The parameter
        \begin{equation}
            \tau_{0} = n_{1}\frac{z_{0}}{c} - t_{0}\,,
        \end{equation}
       denotes the traveling-wave coordinate associated with the space-time corner located at~$\left(z_{0},ct_{0}\right)$ (Fig.~\ref{fig:Canonical_Space-Time_Structures}c). The decomposition in Eq.~\eqref{eq:appendix:ST_Corner_Splitting_Incident_Wave} separates the incident waveform into the portion that reaches the stationary interface and the portion that interacts with the temporal interface. Since the stationary and temporal interface problems were solved in Sec.~\ref{subsec:appendix:Space_Interface} and Sec.~\ref{subsec:appendix:Time_Interface}, respectively, the corresponding solutions can be applied directly using the decomposed incident waves in Eq.~\eqref{eq:appendix:ST_Corner_Splitting_Incident_Wave}. Specifically,~$\psi_{\text{i}}^{\text{S}}$ serves as the incident wave for the stationary interface, while~$\psi_{\text{i}}^{\text{T}}$ serves as the incident wave for the temporal interface. Substituting Eq.~\eqref{eq:appendix:ST_Corner_Splitting_Incident_Wave} into the solutions derived in Eqs.~\eqref{eq:appendix:Space_Interface_Scattered_Waves} and Eqs.~\eqref{eq:appendix:Time_Interface_Scattered_Waves} yields
         \begin{subequations}\label{eq:appendix:ST_Corner_Solutions_First_Scattering_Event}
            \begin{align}
                \psi_{1}^{-\text{S}}{\left[\tau_{1}^{-}\right]} &= \frac{\eta_{2}-\eta_{1}}{\eta_{2}+\eta_{1}}\psi_{\text{i}}^{\text{S}}{\left[-\tau_{1}^{-}+2\frac{z_{0}}{u_{1}}\right]} \nonumber\\
                &= \frac{\eta_{2}-\eta_{1}}{\eta_{2}+\eta_{1}}\psi_{\text{i}}{\left[-\tau_{1}^{-}+2n_{1}\frac{z_{0}}{c}\right]} \nonumber \\
                &\hspace{2cm} \times\Theta{\left[\tau_{0}+\tau_{1}^{-}-2n_{1}\frac{z_{0}}{c}\right]} \,, \\
                \psi_{2}^{+\text{S}}{\left[\tau_{2}^{+}\right]} &= \frac{2\eta_{2}}{\eta_{2}+\eta_{1}}\psi_{\text{i}}^{\text{S}}{\left[\tau_{2}^{+} + \left(n_{1}-n_{2}\right)\frac{z_{0}}{c}\right]} \nonumber \\
                &= \frac{2\eta_{2}}{\eta_{2}+\eta_{1}}\psi_{\text{i}}{\left[\tau_{2}^{+} + \left(n_{1}-n_{2}\right)\frac{z_{0}}{c}\right]} \nonumber \\
                &\hspace{1cm} \times\Theta{\left[\tau_{0}-\tau_{2}^{+} - \left(n_{1}-n_{2}\right)\frac{z_{0}}{c}\right]}\,, \\
                \psi_{2}^{-\text{T}}{\left[\tau_{2}^{-}\right]} &= \frac{n_{1}}{n_{2}}\frac{\eta_{2}-\eta_{1}}{2\eta_{1}}\psi_{\text{i}}^{\text{T}}{\left[\frac{n_{1}}{n_{2}}\tau_{2}^{-} - \left(1+\frac{n_{1}}{n_{2}}\right)t_{0}\right]} \nonumber \\
                &= \frac{n_{1}}{n_{2}}\frac{\eta_{2}-\eta_{1}}{2\eta_{1}}\psi_{\text{i}}{\left[\frac{n_{1}}{n_{2}}\tau_{2}^{-} - \left(1+\frac{n_{1}}{n_{2}}\right)t_{0}\right]} \nonumber \\
                &\hspace{0.5cm} \times\Theta{\left[\frac{n_{1}}{n_{2}}\tau_{2}^{-} - \left(1+\frac{n_{1}}{n_{2}}\right)t_{0} - \tau_{0}\right]} \,, \label{eq:appendix:ST_Corner_Solution_Psi_2_MinT}\\ 
                \psi_{2}^{+\text{T}}{\left[\tau_{2}^{+}\right]} &= \frac{n_{1}}{n_{2}}\frac{\eta_{2}+\eta_{1}}{2\eta_{1}}\psi_{\text{i}}^{\text{T}}{\left[\frac{n_{1}}{n_{2}}\tau_{2}^{-} - \left(1-\frac{n_{1}}{n_{2}}\right)t_{0}\right]} \nonumber \\
                &= \frac{n_{1}}{n_{2}}\frac{\eta_{2}+\eta_{1}}{2\eta_{1}}\psi_{\text{i}}{\left[\frac{n_{1}}{n_{2}}\tau_{2}^{-} - \left(1-\frac{n_{1}}{n_{2}}\right)t_{0}\right]} \nonumber \\
                &\hspace{0.5cm} \times\Theta{\left[\frac{n_{1}}{n_{2}}\tau_{2}^{-} - \left(1-\frac{n_{1}}{n_{2}}\right)t_{0} - \tau_{0}\right]}\,.
            \end{align}
        \end{subequations}

        \pati{Second Scattering Event}{
        }
        
        The second scattering event is produced by the later-backward wave~$\psi_{2}^{-\text{T}}$, which propagates toward the stationary interface. This wave therefore behaves as an incident wave traveling from medium~$2$ toward medium~$1$. Since this interaction is identical to the stationary-interface problem considered in Sec.~\ref{subsec:appendix:Space_Interface}, the corresponding solutions are obtained directly by interchanging the medium indices~$1$ and~$2$ in Eqs.~\eqref{eq:appendix:Space_Interface_Scattered_Waves}. Substituting the expression for~$\psi_{2}^{-\text{T}}$ from Eq.~\eqref{eq:appendix:ST_Corner_Solution_Psi_2_MinT} into Eqs.~\eqref{eq:appendix:Space_Interface_Scattered_Waves} then gives
        \begin{subequations}\label{eq:appendix:ST_Corner_Solutions_Second_Scattering_Event}
            \begin{align}
                \psi_{1}^{-\text{ST}}{\left[\tau_{1}^{-}\right]} &= \frac{2\eta_{1}}{\eta_{1}+\eta_{2}}\psi_{2}^{-\text{T}}{\left[\tau_{1}^{-} + \left(n_{2}-n_{1}\right)\frac{z_{0}}{c}\right]} \nonumber \\
                &= \frac{n_{1}}{n_{2}}\frac{2\eta_{1}\left(\eta_{2}-\eta_{1}\right)}{\left(\eta_{2}+\eta_{1}\right)^{2}} \nonumber \\
                &\times\psi_{\text{i}}{\left[\frac{n_{1}}{n_{2}}\tau_{1}^{-} + \frac{n_{1}}{n_{2}}\left(n_{2}-n_{1}\right)\frac{z_{0}}{c} - \left(1+\frac{n_{1}}{n_{2}}\right)t_{0}\right]} \nonumber \\
                &\times\Theta{\left[\frac{n_{1}}{n_{2}}\tau_{1}^{-} + \frac{n_{1}}{n_{2}}\left(n_{2}-n_{1}\right)\frac{z_{0}}{c} - \left(1+\frac{n_{1}}{n_{2}}\right)t_{0} - \tau_{0}\right]}\,, \\
                \psi_{2}^{+\text{ST}}{\left[\tau_{2}^{+}\right]} &= \frac{\eta_{1}-\eta_{2}}{\eta_{1}+\eta_{2}}\psi_{2}^{-\text{T}}{\left[-\tau_{2}^{+}+2\frac{z_{0}}{u_{2}}\right]} \nonumber \\
                &= -\frac{n_{1}}{n_{2}}\frac{\left(\eta_{2}-\eta_{1}\right)^{2}}{2\eta_{1}\left(\eta_{2}+\eta_{1}\right)} \nonumber \\
                &\times\psi_{\text{i}}{\left[-\frac{n_{1}}{n_{2}}\tau_{2}^{+}+2n_{1}\frac{z_{0}}{c} - \left(1+\frac{n_{1}}{n_{2}}\right)t_{0}\right]}\nonumber \\
                &\times\Theta{\left[-\frac{n_{1}}{n_{2}}\tau_{2}^{+}+2n_{1}\frac{z_{0}}{c} - \left(1+\frac{n_{1}}{n_{2}}\right)t_{0} - \tau_{0}\right]}\,,
            \end{align}
        \end{subequations}
        The total scattered fields are obtained by superposing all contributions according to Eqs.~\eqref{eq:appendix:ST_Corner_Sum_Scattered_Waves}, with the individual waveforms given by Eqs.~\eqref{eq:appendix:ST_Corner_Solutions_First_Scattering_Event} and Eqs.~\eqref{eq:appendix:ST_Corner_Solutions_Second_Scattering_Event}.

    \subsection{Constant-Velocity Traveling Interface}\label{subsec:appendix:Constant-Velocity_Traveling_Interface}
        \pati{Scattered Waves}{
        }

         At a constant-velocity traveling interface in the subluminal regime ($\left|v_{\text{m}}\right| < c/n_{2}$)~\footnote{The other two regimes, the superluminal ($\left|v_{\text{m}}\right| > c/n_{1}$) and interluminal ($c/n_{2} < \left|v_{\text{m}}\right| < c/n_{1}$) regime, involve radically different physics. They are treated in~\cite{DeKinder2026_Scat_Chirp_ASTEM_PUB}.}, the scattered waves are~$\psi_{1}^{-}$ (reflection) and~$\psi_{2}^{+}$ (transmission), see Fig.~\ref{fig:Canonical_Space-Time_Structures}d. The interface is described by~$z = v_{\text{m}}t + z_{0}$. In this case, Eqs.~\eqref{eq:appendix:Boundary_Conditions_Psi} result in
         \begin{widetext}
             \begin{subequations}\label{eq:appendix:Constant_Interface_Boundary_Conditions}
                \begin{align}
                    &\left(1 - n_{1}\frac{v_{\text{m}}}{c}\right)\psi_{\text{i}}{\left[-\left(1-n_{1}\frac{v_{\text{m}}}{c}\right)t^{\star}+n_{1}\frac{z_{0}}{c}\right]} + \left(1 + n_{1}\frac{v_{\text{m}}}{c}\right)\psi_{1}^{-}{\left[\left(1+n_{1}\frac{v_{\text{m}}}{c}\right)t^{\star}+n_{1}\frac{z_{0}}{c}\right]} \nonumber\\
                    &\hspace{8cm}= \left(1 - n_{2}\frac{v_{\text{m}}}{c}\right)\psi_{2}^{+}{\left[-\left(1-n_{2}\frac{v_{\text{m}}}{c}\right)t^{\star}+n_{2}\frac{z_{0}}{c}\right]}\,, \\
                    &\frac{1}{\eta_{1}}\left(1 - n_{1}\frac{v_{\text{m}}}{c}\right)\psi_{\text{i}}{\left[-\left(1-n_{1}\frac{v_{\text{m}}}{c}\right)t^{\star}+n_{1}\frac{z_{0}}{c}\right]} -\frac{1}{\eta_{1}}\left(1 + n_{1}\frac{v_{\text{m}}}{c}\right)\psi_{1}^{-}{\left[\left(1+n_{1}\frac{v_{\text{m}}}{c}\right)t^{\star}+n_{1}\frac{z_{0}}{c}\right]} \nonumber \\
                    &\hspace{8cm}= \frac{1}{\eta_{2}}\left(1 - n_{2}\frac{v_{\text{m}}}{c}\right)\psi_{2}^{+}{\left[-\left(1-n_{2}\frac{v_{\text{m}}}{c}\right)t^{\star}+n_{2}\frac{z_{0}}{c}\right]}\,,
                \end{align}
            \end{subequations}
        \end{widetext}
        where $z{\left[t^{\star}\right]} = z_{\text{m}}t^{\star} + z_{0}$. Solving Eqs.~\eqref{eq:appendix:Constant_Interface_Boundary_Conditions} for $\psi_{1}^{-}$ and $\psi_{2}^{+}$ yields the scattering solutions at the space-time interface,
        \begin{widetext}            
            \begin{subequations}\label{eq:appendix:Constant_Interface_Solution_at_Interface}
                \begin{align}
                    \psi_{1}^{-}{\left[\left(1+n_{1}\frac{v_{\text{m}}}{c}\right)t^{\star}+n_{1}\frac{z_{0}}{c}\right]}  &= \frac{\eta_{2}-\eta_{1}}{\eta_{2}+\eta_{1}}\frac{1-n_{1}v_{\text{m}}/c}{1+n_{1}v_{\text{m}}/c}\psi_{\text{i}}{\left[-\left(1-n_{1}\frac{v_{\text{m}}}{c}\right)t^{\star}+n_{1}\frac{z_{0}}{c}\right]}\,, \\
                    \psi_{2}^{+}{\left[-\left(1-n_{2}\frac{v_{\text{m}}}{c}\right)t^{\star}+n_{2}\frac{z_{0}}{c}\right]} &= \frac{2\eta_{2}}{\eta_{2}+\eta_{1}}\frac{1-n_{1}v_{\text{m}}/c}{1-n_{2}v_{\text{m}}/c}\psi_{\text{i}}{\left[-\left(1-n_{1}\frac{v_{\text{m}}}{c}\right)t^{\star}+n_{1}\frac{z_{0}}{c}\right]}\,.
                \end{align}
            \end{subequations}
        \end{widetext}
        To determine the fields away from the boundary, we have to extend Eqs.~\eqref{eq:appendix:Constant_Interface_Solution_at_Interface} to the entire space-time domain. Equation the pulse trajectory [Eq.~\eqref{eq:appendix:Space_Interface_Trajectory_Pulse}] to the interface trajectory and solving for the scattering time yields
        \begin{equation}\label{eq:appendix:Constant_Interface_Coordinate_Transformation}
            t^{\star} = \mp\frac{\tau_{i}^{\pm}-n_{i}z_{0}/c}{1\mp n_{i}v_{\text{m}}/c}\,.
        \end{equation}
        Inserting Eq.~\eqref{eq:appendix:Constant_Interface_Coordinate_Transformation} into Eqs.~\eqref{eq:appendix:Constant_Interface_Solution_at_Interface} yields the final solution for the entire space-time domain:
        \begin{subequations}\label{eq:appendix:Constant_Interface_Scattered_Waves}
            \begin{align}
                &\psi_{1}^{-}{\left[\tau_{1}^{-}\right]} = \frac{\eta_{2}-\eta_{1}}{\eta_{2}+\eta_{1}}\frac{1-n_{1}v_{\text{m}}/c}{1+n_{1}v_{\text{m}}/c} \nonumber \\
                &\times\psi_{\text{i}}{\left[-\frac{1-n_{1}v_{\text{m}}/c}{1+n_{1}v_{\text{m}}/c}\tau_{1}^{-} + n_{1}\left(1+\frac{1-n_{1}v_{\text{m}}/c}{1+n_{1}v_{\text{m}}/c}\right)\frac{z_{0}}{c}\right]}\,, \\
                &\psi_{2}^{+}{\left[\tau_{2}^{+}\right]} = \frac{2\eta_{2}}{\eta_{2}+\eta_{1}}\frac{1-n_{1}v_{\text{m}}/c}{1-n_{2}v_{\text{m}}/c} \nonumber \\
                &\times\psi_{\text{i}}{\left[\frac{1-n_{1}v_{\text{m}}/c}{1-n_{2}v_{\text{m}}/c}\tau_{2}^{+} + \left(n_{1}-n_{2}\frac{1-n_{1}v_{\text{m}}/c}{1-n_{2}v_{\text{m}}/c}\right)\frac{z_{0}}{c}\right]}\,.
            \end{align}
        \end{subequations}

    \subsection{Space-Time Wedges}\label{subsec:appendix:Space-Time_Wedges}
    \pati{Scattered Waves}{
    }
    
    This canonical space-time structure example consists of two interfaces moving at different constant velocities, see Fig.~\ref{fig:Canonical_Space-Time_Structures}e~\cite{Bahrami2025_Wedges_USTEM_PUB}. The interfaces are parametrized as~$z=v_{\text{m},i}t + z_{i}$, where~$v_{\text{m},i}$ denotes the modulation velocity and~$z_{i}$ the initial position of interface~$i = 1, 2$. The waves inside the wedge are denoted by~$\psi_{2}^{+}$ and~$\psi_{2}^{-}$, representing the forward and backward propagating waves, respectively.    
    \par
    As there are two interfaces, four boundary conditions must be applied, with two conditions imposed at each interface. Application of Eqs.~\eqref{eq:appendix:Boundary_Conditions_Psi} to the space-time wedge gives the following boundary conditions at the first interface:
    \begin{widetext}
        \begin{subequations}\label{eq:appendix:ST_Wedge_Boundary_Condition_First_Interface}
            \begin{align}
                &M_{11}^{-}\psi_{\text{i}}{\left[-M_{11}^{-}t^{\star}+n_{1}\frac{z_{1}}{c}\right]} + M_{11}^{+}\psi_{1}^{-}{\left[M_{11}^{+}t^{\star}+n_{1}\frac{z_{1}}{c}\right]} \nonumber \\
                &\hspace{7cm}= M_{12}^{-}\psi_{2}^{+}{\left[-M_{12}^{-}t^{\star}+n_{2}\frac{z_{1}}{c}\right]} + M_{12}^{+}\psi_{2}^{-}{\left[M_{12}^{+}t^{\star}+n_{2}\frac{z_{1}}{c}\right]}\,, \\
                &\frac{1}{\eta_{1}}M_{11}^{-}\psi_{\text{i}}{\left[-M_{11}^{-}t^{\star}+n_{1}\frac{z_{1}}{c}\right]} - \frac{1}{\eta_{1}}M_{11}^{+}\psi_{1}^{-}{\left[M_{11}^{+}t^{\star}+n_{1}\frac{z_{1}}{c}\right]} \nonumber \\
                &\hspace{7cm}= \frac{1}{\eta_{2}}M_{12}^{-}\psi_{2}^{+}{\left[-M_{12}^{-}t^{\star}+n_{2}\frac{z_{1}}{c}\right]} - \frac{1}{\eta_{2}}M_{12}^{+}\psi_{2}^{-}{\left[M_{12}^{+}t^{\star}+n_{2}\frac{z_{1}}{c}\right]} \,,
            \end{align}
        \end{subequations}
    \end{widetext}
    where 
    \begin{equation}
        M^{\pm}_{ij}=1\pm n_{j}\frac{v_{\text{m},i}}{c}\,.
    \end{equation}
    For the second interface, the moving boundary conditions are
    \begin{widetext}        
        \begin{subequations}\label{eq:appendix:ST_Wedge_Boundary_Condition_Second_Interface}
            \begin{align}
                M_{22}^{-}\psi_{2}^{+}{\left[-M_{22}^{-}t^{\star}+n_{2}\frac{z_{2}}{c}\right]} + M_{22}^{+}\psi_{2}^{-}{\left[M_{22}^{+}t^{\star}+n_{2}\frac{z_{2}}{c}\right]} &= M_{21}^{-}\psi_{1}^{+}{\left[-M_{21}^{-}t^{\star}+n_{1}\frac{z_{2}}{c}\right]} \,, \\
                \frac{1}{\eta_{2}}M_{22}^{-}\psi_{2}^{+}{\left[-M_{22}^{-}t^{\star}+n_{2}\frac{z_{2}}{c}\right]} - \frac{1}{\eta_{2}}M_{22}^{+}\psi_{2}^{-}{\left[M_{22}^{+}t^{\star}+n_{2}\frac{z_{2}}{c}\right]} &= \frac{1}{\eta_{1}}M_{21}^{-}\psi_{1}^{+}{\left[-M_{21}^{-}t^{\star}+n_{1}\frac{z_{2}}{c}\right]} \,.
            \end{align}
        \end{subequations}
    \end{widetext}
    Next, we eliminate $\psi_{1}^{-}$ from Eqs.~\eqref{eq:appendix:ST_Wedge_Boundary_Condition_First_Interface} and $\psi_{1}^{+}$ from Eq.~\eqref{eq:appendix:ST_Wedge_Boundary_Condition_Second_Interface} to arrive at the two equations
    \begin{widetext}        
        \begin{subequations}\label{eq:appendix:ST_Wedge_BC_TEMP}
            \begin{align}
                M_{12}^{-}\psi_{2}^{+}{\left[-M_{12}^{-}t^{\star}+n_{2}\frac{z_{1}}{c}\right]} + RM_{12}^{+}\psi_{2}^{-}{\left[M_{12}^{+}t^{\star}+n_{2}\frac{z_{1}}{c}\right]} &= T_{12}M_{11}^{-}\psi_{\text{i}}{\left[-M_{11}^{-}t^{\star} + n_{1}\frac{z_{1}}{c}\right]}\,, \label{eq:appendix:ST_Wedge_BC_TEMP_1}\\
                RM_{22}^{-}\psi_{2}^{+}{\left[-M_{22}^{-}t^{\star}+n_{2}\frac{z_{2}}{c}\right]} + M_{22}^{+}\psi_{2}^{-}{\left[M_{22}^{+}t^{\star}+n_{2}\frac{z_{2}}{c}\right]} &= 0\,, \label{eq:appendix:ST_Wedge_BC_TEMP_2}
            \end{align}
        \end{subequations}
    \end{widetext}
    where 
    \begin{subequations}
        \begin{align}
            R &= \frac{\eta_{2}-\eta_{1}}{\eta_{2}+\eta_{1}}\,, \\
            T_{ij} &= \frac{2\eta_{j}}{\eta_{i}+\eta_{j}}\,,
        \end{align}
    \end{subequations}
    are the stationary reflection and transmission coefficients, respectively. Noticing that the arguments of~$\psi_{2}^{-}$ in Eqs.~\eqref{eq:appendix:ST_Wedge_BC_TEMP} are different, we perform in Eq.~\eqref{eq:appendix:ST_Wedge_BC_TEMP_2} the change of variable
    \begin{equation}\label{eq:ST_Wedge_TEMP_Change_of_Variable}
        t^{\star} \mapsto \frac{M_{12}^{+}}{M_{22}^{+}}t^{\star} + \frac{n_{2}}{M_{22}^{+}}\frac{z_{1}-z_{2}}{c} \,,
    \end{equation}
    which yields upon substitution into Eq.~\eqref{eq:appendix:ST_Wedge_BC_TEMP_2}
    \begin{equation}\label{eq:appendix:ST_Wedge_BC_TEMP_3}
        \begin{split}
            &\psi_{2}^{-}{\left[M_{12}^{+}t^{\star} + n_{2}\frac{z_{1}}{c}\right]} = -R\frac{M_{22}^{-}}{M_{22}^{+}} \\
            &\hspace{0.1cm}\times\psi_{2}^{+}{\left[-\frac{M_{22}^{-}M_{12}^{+}}{M_{22}^{+}}t^{\star} + n_{2}\frac{z_{2}}{c} - n_{2}\frac{M_{22}^{-}}{M_{22}^{+}}\frac{z_{1}-z_{2}}{c}\right]}\,,
        \end{split}
    \end{equation}
    such that the argument of $\psi_{2}^{-}$ is the same as in Eq.~\eqref{eq:appendix:ST_Wedge_BC_TEMP_1}. Inserting Eq.~\eqref{eq:appendix:ST_Wedge_BC_TEMP_3} into Eq.~\eqref{eq:appendix:ST_Wedge_BC_TEMP_1} gives then
    \begin{equation}
        \begin{split}
            &\psi_{2}^{+}{\left[-M_{12}^{-}t^{\star} + n_{2}\frac{z_{1}}{c}\right]} = T_{12}\frac{M_{11}^{-}}{M_{12}^{-}}\psi_{\text{i}}{\left[-M_{11}^{-}t^{\star} + n_{1}\frac{z_{1}}{c}\right]} \\
            &+R^{2}\frac{M_{12}^{+}M_{22}^{-}}{M_{12}^{-}M_{22}^{+}}\psi_{2}^{+}{\left[-\frac{M_{22}^{-}M_{12}^{+}}{M_{22}^{+}}t^{\star} + n_{2}\frac{z_{2}}{c} - n_{2}\frac{M_{22}^{-}}{M_{22}^{+}}\frac{z_{1}-z_{2}}{c}\right]}
        \end{split}
    \end{equation}
    We now easily solve this equation by writing in terms of $x= - M_{12}^{-}t^{\star} + n_{2}z_{1}/c$ as
    \begin{equation}\label{eq:appendix:ST_Wedge_Equation_to_Solve_Psi_2_Plus}
        \psi_{2}^{+}{\left[x\right]} = X\psi_{\text{i}}{\left[A+Bx\right]} + Y\psi_{2}^{+}{\left[C+Dx\right]}\,, 
    \end{equation}
    where
    \begin{subequations}
        \begin{align}
            X &= T_{12}\frac{M_{11}^{-}}{M_{12}^{-}}\,, \\
            A &= n_{1}\frac{z_{1}}{c} - n_{2}\frac{M_{11}^{-}}{M_{12}^{-}}\frac{z_{1}}{c}\,, \\
            B &= \frac{M_{11}^{-}}{M_{12}^{-}}\,, \\
            Y &= R^{2}\frac{M_{12}^{+}M_{22}^{-}}{M_{12}^{-}M_{22}^{+}}\,, \\
            C &= n_{2}\frac{z_{2}}{c} - n_{2}\frac{M_{22}^{-}}{M_{22}^{+}}\frac{z_{1}-z_{2}}{c} - n_{2}\frac{M_{12}^{+}M_{22}^{-}}{M_{12}^{-}M_{22}^{+}}\frac{z_{1}}{c}\,, \\
            D &= \frac{M_{12}^{+}M_{22}^{-}}{M_{12}^{-}M_{22}^{+}}\,.
        \end{align}
    \end{subequations}
    The recursive structure of Eq.~\eqref{eq:appendix:ST_Wedge_Equation_to_Solve_Psi_2_Plus} allows the solution to be expressed as an infinite series obtained by iteratively evaluating the second term on the right-hand side. This yields the explicit iterative solution
    \begin{equation}\label{eq:appendix:ST_Wedge_Solution_Psi_2_Plus}
        \psi_{2}^{+}{\left[x\right]} = X\sum_{p=0}^{\infty}Y^{p}\psi_{\text{i}}{\left[A+BC\sum_{p'=1}^{p}D^{p'-1} + BD^{p}x\right]}\,.
    \end{equation}
    Next, we eliminate $\psi_{2}^{-}$ from Eqs.~\eqref{eq:appendix:ST_Wedge_Boundary_Condition_Second_Interface} to arrive at the following equation, expressing $\psi_{1}^{+}$ in terms of $\psi_{2}^{+}$:
    \begin{equation}\label{eq:appendix:ST_Wedge_Resolution_Psi_1_Plus}
        \psi_{1}^{+}{\left[-M_{21}^{-}t^{\star} + n_{1}\frac{z_{2}}{c}\right]} = T_{21}\frac{M_{22}^{-}}{M_{21}^{-}}\psi_{2}^{+}{\left[-M_{22}^{-}t^{\star} + n_{2}\frac{z_{2}}{c}\right]}\,.
    \end{equation}
    Next, we apply a similar procedure [Eq.~\eqref{eq:ST_Wedge_TEMP_Change_of_Variable} to Eq.~\eqref{eq:appendix:ST_Wedge_Resolution_Psi_1_Plus}] to determine $\psi_{2}^{-}$:
    \begin{equation}
        \psi_{2}^{-}{\left[x\right]} = X'\psi_{\text{i}}{\left[A'+B'x\right]} + Y'\psi_{2}^{-}{\left[C'+D'x\right]}\,,
    \end{equation}
    where
    \begin{subequations}
        \begin{align}
            X' &= -RT_{12}\frac{M_{11}^{-}M_{22}^{-}}{M_{12}^{-}M_{22}^{+}}\, \\
            A' &= n_{1}\frac{z_{1}}{c} + n_{2}\frac{M_{11}^{-}M_{22}^{-}}{M_{12}^{-}M_{22}^{+}}\frac{z_{2}}{c} + n_{2}\frac{M_{11}^{-}}{M_{12}^{-}}\frac{z_{2}-z_{1}}{c}\,, \\
            B' &= -\frac{M_{11}^{-}M_{22}^{-}}{M_{12}^{-}M_{22}^{+}}\,, \\
            Y' &= R^{2}\frac{M_{12}^{+}M_{22}^{-}}{M_{12}^{-}M_{22}^{+}}\,, \\
            C' &= n_{2}\frac{z_{1}}{c}-n_{2}\frac{M_{12}^{+}M_{22}^{-}}{M_{12}^{-}M_{22}^{+}}\frac{z_{2}}{c} -n_{2}\frac{M_{12}^{+}}{M_{12}^{-}}\frac{z_{2}-z_{1}}{c}\,, \\
            D' &= \frac{M_{12}^{+}M_{22}^{-}}{M_{12}^{-}M_{22}^{+}}\,,
        \end{align}
    \end{subequations}
    whose resolution yields $\psi_{2}^{-}$,
    \begin{equation}\label{eq:appendix:ST_Wedge_Solution_Psi_2_Min}
        \psi_{2}^{-}{\left[x\right]} = X'\sum_{p=0}^{\infty}Y'^{p}\psi_{\text{i}}{\left[A'+B'C'\sum_{p'=1}^{p}D'^{p'-1}+B'D'^{p}x\right]}\,.
    \end{equation}
    By eliminating $\psi_{2}^{+}$ in Eqs.~\eqref{eq:appendix:ST_Wedge_Boundary_Condition_First_Interface}, we can express $\psi_{1}^{-}$ in terms of $\psi_{2}^{-}$ (and $\psi_{\text{i}}$), viz.,
    \begin{align}\label{eq:ST_Wedge_Resolution_Psi_2_Min}
            &\psi_{1}^{-}{\left[M_{11}^{+}t^{\star} + n_{1}\frac{z_{1}}{c}\right]} = R\frac{M_{11}^{-}}{M_{11}^{+}}\psi_{\text{i}}{\left[-M_{11}^{-}t^{\star}+n_{1}\frac{z_{1}}{c}\right]} \nonumber \\
            &\hspace{1cm}+ T_{21}\frac{M_{12}^{+}}{M_{11}^{+}}\psi_{2}^{-}{\left[M_{12}^{+}t^{\star}+n_{2}\frac{z_{1}}{c}\right]}\,. 
    \end{align}
    At this point, we have $\psi_{1}^{+}$ and $\psi_{1}^{-}$ in Eq.~\eqref{eq:appendix:ST_Wedge_Resolution_Psi_1_Plus} and Eq.~\eqref{eq:ST_Wedge_Resolution_Psi_2_Min}, as functions of $\psi_{2}^{\pm}$, respectively. The arguments of $\psi_{1}^{\pm}$ do not have the standard traveling wave argument [Eq.~\eqref{eq:appendix:Traveling_Wave_Variables}], but they can be forced to the initial traveling-wave forms, by applying the STESEM procedure, yielding
    \begin{subequations}\label{eq:appendix:ST_Wedge_TEMP_Final_Solutions}
        \begin{align}
            \psi_{1}^{+}{\left[\tau_{1}^{+}\right]} &= T_{21}\frac{M_{22}^{-}}{M_{21}^{-}}\psi_{2}^{+}{\left[\frac{M_{22}^{-}}{M_{21}^{-}}\tau_{1}^{+} + \left(n_{2}-n_{1}\frac{M_{22}^{-}}{M_{21}^{-}}\right)\frac{z_{2}}{c}\right]}\,, \\
            \psi_{1}^{-}{\left[\tau_{1}^{-}\right]} &= R\frac{M_{11}^{-}}{M_{11}^{+}}\psi_{\text{i}}{\left[-\frac{M_{11}^{-}}{M_{11}^{+}}\tau_{1}^{-} + n_{1}\left(1+\frac{M_{11}^{-}}{M_{11}^{+}}\right)\frac{z_{1}}{c}\right]} \nonumber \\
            &+T_{21}\frac{M_{12}^{+}}{M_{11}^{+}}\psi_{2}^{-}{\left[\frac{M_{12}^{+}}{M_{11}^{+}}\tau_{1}^{-} + \left(n_{2}-n_{1}\frac{M_{12}^{+}}{M_{11}^{+}}\right)\frac{z_{1}}{c}\right]}\,.
        \end{align}
    \end{subequations}
    These equations can be written in final forms by substituting Eq.~\eqref{eq:appendix:ST_Wedge_Solution_Psi_2_Plus} and Eq.~\eqref{eq:appendix:ST_Wedge_Solution_Psi_2_Min} into Eqs.~\eqref{eq:appendix:ST_Wedge_TEMP_Final_Solutions}, which yields
    \begin{subequations}
        \begin{align}
            \psi_{2}^{+}{\left[\tau_{2}^{+}\right]} &= T_{21}\frac{M_{22}^{-}}{M_{21}^{-}}X\sum_{p=0}^{\infty}Y^{p}\psi_{\text{i}}{\left[BD^{p}\frac{M_{22}^{-}}{M_{21}^{-}}\tau_{2}^{+} + \Delta\tau_{p}\right]}\,, \\
            \psi_{1}^{-}{\left[\tau_{1}^{-}\right]} &= R\frac{M_{11}^{-}}{M_{11}^{+}}\psi_{\text{i}}{\left[-\frac{M_{11}^{-}}{M_{11}^{+}}\tau_{1}^{-} + n_{1}\left(1+n_{1}\frac{M_{11}^{-}}{M_{11}^{+}}\right)\frac{z_{1}}{c}\right]} \nonumber \\
            &T_{21}\frac{M_{12}^{+}}{M_{11}^{+}}X'\sum_{p=0}^{\infty}Y'^{p}\psi_{\text{i}}{\left[B'D'^{p}\frac{M_{12}^{+}}{M_{11}^{+}}\tau_{1}^{-} + \Delta\tau'_{p}\right]}\,,
        \end{align}
    \end{subequations}
    where
    \begin{subequations}
        \begin{align}
            \Delta\tau_{p} &= A + BC\sum_{p'=1}^{p}D^{p'-1} + BD^{p}\left(n_{2}-n_{1}\frac{M_{22}^{-}}{M_{21}^{-}}\right)\frac{z_{2}}{c}\,, \\
            \Delta\tau'_{p} &= A'+ B'C'\sum_{p'=1}^{p}D'^{p'-1}+B'D'^{p}\left(n_{2} - n_{1}\frac{M_{12}^{+}}{M_{11}^{+}}\right)\frac{z_{1}}{c}\,.
        \end{align}
    \end{subequations}
    These equations can be more conveniently written as
    \begin{subequations}
        \begin{align}
            \psi_{2}^{+}{\left[\tau_{2}^{+}\right]} &= T_{12}T_{21}H\sum_{p=0}^{\infty}R^{2p}D^{p}\psi_{\text{i}}{\left[HD^{p}\tau_{2}^{+} + \Delta\tau_{p}\right]}\,, \\
            \psi_{1}^{-}{\left[\tau_{1}^{-}\right]} &= R\frac{M_{11}^{-}}{M_{11}^{+}}\psi_{\text{i}}{\left[-\frac{M_{11}^{-}}{M_{11}^{+}}\tau_{1}^{-} + \tau'_{0}\right]} \nonumber \\
            &- T_{12}T_{21}H'\sum_{p=0}^{\infty}R^{2p+1}D'^{p}\psi_{\text{i}}{\left[H'D'^{p}\tau_{1}^{-} + \Delta\tau'_{p}\right]}\,, 
        \end{align}
    \end{subequations}
    where
    \begin{subequations}
        \begin{align}
            \Delta\tau_{p} &= A + BC\frac{1-D^{p}}{1-D} + BD^{p}\left(n_{2} - n_{1}\frac{M_{22}^{-}}{M_{21}^{-}}\right)\frac{z_{2}}{c}\,, \\
            \Delta\tau'_{0} &= n_{1}\left(1+\frac{M_{11}^{-}}{M_{11}^{+}}\right)\frac{z_{1}}{c}\,, \\
            \Delta\tau'_{p} &= A'+B'C'\frac{1-D'^{p}}{1-D'} + B'D'^{p}\left(n_{2} - n_{1}\frac{M_{12}^{+}}{M_{11}^{+}}\right)\frac{z_{1}}{c}\,,
        \end{align}
    \end{subequations}
    and
    \begin{subequations}
        \begin{align}
            H &= \frac{M_{22}^{-}M_{11}^{-}}{M_{21}^{-}M_{12}^{-}}\,, \\
            H' &= -\frac{M_{12}^{+}M_{11}^{-}M_{22}^{-}}{M_{11}^{+}M_{12}^{-}M_{22}^{+}}\,.
        \end{align}
    \end{subequations}

    \subsection{Accelerated Interface}\label{subsec:appendix:Accelerated_Interface}
        \pati{Scattered Waves}{
        }

        At an accelerated subluminal interface, the scattered waves are~$\psi_{1}^{-}$ and~$\psi_{2}^{+}$, see Fig.~\ref{fig:Canonical_Space-Time_Structures}f~\cite{DeKinder2026_Scat_Chirp_ASTEM_PUB}. The interface is parametrized in a general form as~$z=z{\left[t\right]}$. Application of Eqs.~\eqref{eq:appendix:Boundary_Conditions_Psi} for an accelerated interface yields
        \begin{widetext}            
            \begin{subequations}\label{eq:appendix:Accelerated_Interface_Boundary_Conditions}
                \begin{align}
                    &\left(1 - n_{1}\frac{v_{\text{m}}{\left[t^{\star}\right]}}{c}\right)\psi_{\text{i}}{\left[n_{1}\frac{z{\left[t^{\star}\right]}}{c}-t^{\star}\right]} + \left(1 + n_{1}\frac{v_{\text{m}}{\left[t^{\star}\right]}}{c}\right)\psi_{1}^{-}{\left[n_{1}\frac{z{\left[t^{\star}\right]}}{c}+t^{\star}\right]} = \left(1 - n_{2}\frac{v_{\text{m}}{\left[t^{\star}\right]}}{c}\right)\psi_{2}^{+}{\left[n_{2}\frac{z{\left[t^{\star}\right]}}{c}-t^{\star}\right]}\,, \\
                    &\frac{1}{\eta_{1}}\left(1 - n_{1}\frac{v_{\text{m}}{\left[t^{\star}\right]}}{c}\right)\psi_{\text{i}}{\left[n_{1}\frac{z{\left[t^{\star}\right]}}{c}-t^{\star}\right]} -\frac{1}{\eta_{1}}\left(1 + n_{1}\frac{v_{\text{m}}{\left[t^{\star}\right]}}{c}\right)\psi_{1}^{-}{\left[n_{1}\frac{z{\left[t^{\star}\right]}}{c}+t^{\star}\right]} = \frac{1}{\eta_{2}}\left(1 - n_{2}\frac{v_{\text{m}}{\left[t^{\star}\right]}}{c}\right)\psi_{2}^{+}{\left[n_{2}\frac{z{\left[t^{\star}\right]}}{c}-t^{\star}\right]}\,,
                \end{align}
            \end{subequations}
        \end{widetext}
        which can be solved for~$\psi_{1}^{-}$ and~$\psi_{2}^{+}$ at the interface:
        \begin{subequations}\label{eq:appendix:Accelerated_Interface_Solution_at_Interface}
            \begin{align}
                &\psi_{1}^{-}{\left[n_{1}\frac{z{\left[t^{\star}\right]}}{c}+t^{\star}\right]} = \frac{\eta_{2}-\eta_{1}}{\eta_{2}+\eta_{1}}\frac{1-n_{1}v_{\text{m}}{\left[t^{\star}\right]}/c}{1+n_{1}v_{\text{m}}{\left[t^{\star}\right]}/c} \nonumber \\
                &\hspace{4cm} \times\psi_{\text{i}}{\left[n_{1}\frac{z{\left[t^{\star}\right]}}{c}-t^{\star}\right]}\,, \\
                &\psi_{2}^{+}{\left[n_{2}\frac{z{\left[t^{\star}\right]}}{c}-t^{\star}\right]} = \frac{2\eta_{2}}{\eta_{2}+\eta_{1}}\frac{1-n_{1}v_{\text{m}}{\left[t^{\star}\right]}/c}{1-n_{2}v_{\text{m}}{\left[t^{\star}\right]}/c} \nonumber \\
                &\hspace{4cm} \times\psi_{\text{i}}{\left[n_{1}\frac{z{\left[t^{\star}\right]}}{c}-t^{\star}\right]}\,.
            \end{align}
        \end{subequations}
        To determine the fields away from the interface, we extend Eqs.~\eqref{eq:appendix:Accelerated_Interface_Solution_at_Interface} to the entire space-time domain. Equating the pulse trajectory [Eq.~\eqref{eq:appendix:Space_Interface_Trajectory_Pulse}] to the interface trajectory~$z{\left[t\right]}$, we find the scattering time associated with the arbitrary space-time point,
        \begin{equation}
            \pm \frac{c}{n_{i}}t^{\star} + \frac{c}{n_{i}}\tau_{i}^{\pm} = z{\left[t^{\star}\right]}\,,
        \end{equation}
        which may be rewritten as
        \begin{equation}\label{eq:appendix:Accelerated_Interface_Coordinate_Transformation_Intermediate}
             n_{i}\frac{z{\left[t^{\star}\right]}}{c} \mp t^{\star}=\tau_{i}^{\pm}\,.
        \end{equation}
        In order to find the scattering time, we define an auxiliary function, $f_{i}^{\pm}$, as
        \begin{equation}\label{eq:appendix:Accelerated_Interface_Definition_Auxiliary_Functions}
            f_{i}^{\pm}{\left[t^{\star}\right]} = n_{i}\frac{z{\left[t^{\star}\right]}}{c}\mp t^{\star}\,,
        \end{equation}
        such that $f_{i}^{\pm}$ is exactly the left-hand-side in Eq.~\eqref{eq:appendix:Accelerated_Interface_Coordinate_Transformation_Intermediate}:
        \begin{equation}
            f_{i}^{\pm}{\left[t^{\star}\right]} = \tau_{i}^{\pm}\,,
        \end{equation}
        which can then be inverted to
        \begin{equation}\label{eq:appendix:Accelerated_Interface_Coordinate_Transformation}
            t^{\star} = \left(f_{i}^{\pm}\right)^{-1}{\left[\tau_{i}^{\pm}\right]}\,.
        \end{equation}
        Equation~\eqref{eq:appendix:Accelerated_Interface_Coordinate_Transformation} maps every observation point onto its unique scattering event for an arbitrary accelerated interface. Inserting Eq.~\eqref{eq:appendix:Accelerated_Interface_Coordinate_Transformation} into Eqs.~\eqref{eq:appendix:Accelerated_Interface_Solution_at_Interface} finally yields the field solutions
        \begin{subequations}\label{eq:appendix:Accelerated_Interface_Scattered_Waves}
            \begin{align}
                \psi_{1}^{-}{\left[\tau_{1}^{-}\right]} &= \frac{\eta_{2}-\eta_{1}}{\eta_{2}+\eta_{1}}\frac{1-n_{1}v_{\text{m}}{\left[\left(f_{1}^{-}\right)^{-1}{\left[\tau_{1}^{-}\right]}\right]}/c}{1+n_{1}v_{\text{m}}{\left[\left(f_{1}^{-}\right)^{-1}{\left[\tau_{1}^{-}\right]}\right]}/c} \nonumber \\
                &\hspace{2cm} \times\psi_{\text{i}}{\left[f_{1}^{+}{\left[\left(f_{1}^{-}\right)^{-1}{\left[\tau_{1}^{-}\right]}\right]}\right]}\,, \\
                \psi_{2}^{+}{\left[\tau_{2}^{+}\right]} &= \frac{2\eta_{2}}{\eta_{2}+\eta_{1}}\frac{1-n_{1}v_{\text{m}}{\left[\left(f_{2}^{+}\right)^{-1}{\left[\tau_{2}^{+}\right]}\right]}/c}{1-n_{2}v_{\text{m}}{\left[\left(f_{2}^{+}\right)^{-1}{\left[\tau_{2}^{+}\right]}\right]}/c} \nonumber \\
                &\hspace{2cm} \times \psi_{\text{i}}{\left[f_{1}^{+}{\left[\left(f_{2}^{+}\right)^{-1}{\left[\tau_{2}^{+}\right]}\right]}\right]}\,.
            \end{align}
        \end{subequations}

        \pati{Doppler Frequency Shifts}{}
        
        The Doppler-induced frequency shift may be obtained by differentiating the phase argument of the scattered waves in Eqs.~\eqref{eq:appendix:Accelerated_Interface_Scattered_Waves} with respect to time. Since both scattered waves have a phase of the form $f_{1}^{+}{\left[t^{\star}\right]}$, with $t^{\star} = \left(f_{i}^{\pm}\right)^{-1}{\left[\phi_{i}^{\pm}\right]}$ [Eq.~\eqref{eq:appendix:Accelerated_Interface_Coordinate_Transformation}], we find, using the definition of $f_{1}^{+}$ in Eq.~\eqref{eq:appendix:Accelerated_Interface_Coordinate_Transformation}:
        \begin{equation}\label{eq:appendix:Accelerated_Interface_Doppler_Shift_First_Result_Doppler}
            \begin{split}
                \frac{\omega_{i}^{\pm}{\left[t^{\star}\right]}}{\omega_{\text{i}}} &= -\frac{\partial}{\partial t}\left(f_{1}^{+}{\left[t^{\star}\right]}\right)  \\
                &= \left(1- \frac{n_{1}}{c}\frac{\dd{z{\left[t^{\star}\right]}}}{\dd{t^{\star}}}\right)\frac{\partial t^{\star}}{\partial t}  \\
                &= \left(1- n_{1}\frac{v_{\text{m}}{\left[t^{\star}\right]}}{c}\right)\frac{\partial t^{\star}}{\partial t} \,. 
            \end{split}
        \end{equation}
        Using the identity $\partial_{x}\left(g^{-1}{\left[x\right]}\right) = 1/g'{\left[g^{-1}{\left[x\right]}\right]}$, we find for the last factor in Eq.~\eqref{eq:appendix:Accelerated_Interface_Doppler_Shift_First_Result_Doppler}:
        \begin{equation} \label{eq:appendix:Accelerated_Interface_Doppler_Shift_Intermediate_Result_Partial_Derivative_tstar}
            \begin{split}
                \frac{\partial t^{\star}}{\partial t} &= \frac{\partial}{\partial t}\left(\left(f_{i}^{\pm}\right)^{-1}{\left[\phi_{i}^{\pm}\right]}\right)  \\
                &= \mp\frac{1}{\partial_{t^{\star}}\left(f_{i}^{\pm}{\left[t^{\star}\right]}\right)}  \\
                &= \mp\frac{1}{n_{i}v_{\text{m}}{\left[t^{\star}\right]}/c \mp 1}\,.
                \end{split}
        \end{equation}
        Inserting Eq.~\eqref{eq:appendix:Accelerated_Interface_Doppler_Shift_Intermediate_Result_Partial_Derivative_tstar} into Eq.~\eqref{eq:appendix:Accelerated_Interface_Doppler_Shift_First_Result_Doppler} yields the frequency shift
        \begin{equation}\label{eq:appendix:Doppler_Shift}
            \frac{\omega_{i}^{\pm}{\left[t^{\star}\right]}}{\omega_{\text{i}}} = \frac{n_{1}v_{\text{m}}{\left[t^{\star}\right]}/c - 1}{n_{i}v_{\text{m}}{\left[t^{\star}\right]}/c \mp 1}\,,
        \end{equation}
        where $t^{\star} = \left(f_{i}^{\pm}\right)^{-1}{\left[\phi_{i}^{\pm}\right]}$ [Eq.~\eqref{eq:appendix:Accelerated_Interface_Coordinate_Transformation}].

\bibliography{STESEM}

\end{document}